\documentclass[journal]{IEEEtran}
\usepackage{cite}
\usepackage{amsmath,amssymb,amsfonts}
\usepackage{algorithm}  
\usepackage{algpseudocode}  
\usepackage{algpseudocode}
\usepackage{graphicx}
\usepackage{epstopdf}
\usepackage{textcomp}
\usepackage{xcolor}
\usepackage{bm}
\usepackage{setspace}

\ifCLASSINFOpdf
\else
\fi
\begin{document}
	
%
\title{Robust Sum-Rate Maximization in Transmissive RMS Transceiver-Enabled SWIPT Networks}
%
%
%
\author{Zhendong~Li,~Wen~Chen,~\IEEEmembership{Senior~Member,~IEEE},~Ziheng~Zhang,~Qingqing~Wu,~\IEEEmembership{Senior~Member,~IEEE},\\~Huanqing~Cao,~and~Jun~Li,~\IEEEmembership{Senior~Member,~IEEE}
\thanks{This work is supported by National key project 2020YFB1807700 and 2018YFB1801102, by Shanghai Kewei 20JC1416502 and 22JC1404000, Pudong PKX2021-D02 and NSFC 62071296.}
\thanks{Z. Li, W. Chen, Z. Zhang, Q. Wu, and H. Cao are with the Department of Electronic Engineering, Shanghai Jiao Tong University, Shanghai 200240, China (e-mail: lizhendong@sjtu.edu.cn; wenchen@sjtu.edu.cn; zhangziheng@sjtu.edu.cn; caohuanqing@sjtu.edu.cn; wu.qq1010@gmail.com).}
\thanks{J. Li is with the School of Electronic and Optical Engineering, Nanjing University of Science Technology, Nanjing 210094, China (email: jun.li@njust.edu.cn). }
\thanks{(\emph{Corresponding author: Wen Chen.})}}

\maketitle

\begin{abstract}
In this paper, we propose a state-of-the-art downlink communication transceiver design for transmissive reconfigurable metasurface (RMS)-enabled simultaneous wireless information and power transfer (SWIPT) networks. Specifically, a feed antenna is deployed in the transmissive RMS-based transceiver, which can be used to implement beamforming. According to the relationship between wavelength and propagation distance, the spatial propagation models of plane and spherical waves are built. Then, in the case of imperfect channel state information (CSI), we formulate a robust system sum-rate maximization problem that jointly optimizes RMS transmissive coefficient, transmit power allocation, and power splitting ratio design while taking account of the non-linear energy harvesting model and outage probability criterion. Since the coupling of optimization variables, the whole optimization problem is non-convex and cannot be solved directly. Therefore, the alternating optimization (AO) framework is implemented to decompose the non-convex original problem. In detail, the whole problem is divided into three sub-problems to solve. For the non-convexity of the objective function, successive convex approximation (SCA) is used to transform it, and penalty function method and difference-of-convex (DC) programming are applied to deal with the non-convex constraints. Finally, we alternately solve the three sub-problems until the entire optimization problem converges. Numerical results show that our proposed algorithm has convergence and better performance than other benchmark algorithms.
\end{abstract}


%
\IEEEpeerreviewmaketitle

\section{Introduction}
\IEEEPARstart{T}{he} rapid development of wireless communication enables the Internet-of-Things (IoT) to be utilized in more scenarios, e.g., smart industry, smart medical and the Internet of vehicles \cite{8088539,9355403}. Based on the relevant data, it is inferred that the number of IoT devices worldwide will rise to 14.7 billion by 2030 in the future IoT networks \cite{9442810}. However, IoT devices are usually small in size, which makes the battery capacity often limited and have difficulty in meeting the energy requirements of rich applications in IoT. Therefore, energy management for large-scale IoT devices is a critical issue. Meanwhile, to solve the path loss problem caused by high-frequency communication and ensure that the coverage is not reduced, the number of 5G base stations (BSs) is greatly increased compared with 4G BSs\cite{8187650}. In addition, massive multiple-input multiple-output (MIMO) requires numerous radio frequency (RF) links to provide support, which will lead to a surge in power consumption and cost. Hence, it is urgent to seek a novel transceiver architecture with low power consumption and low cost.

As a promising technique for energy harvested in the IoT, wireless energy transmission (WET) can convert the received RF signal into electrical energies, which can be well applied to solve the energy management of large-scale IoT devices \cite{9625847}. Simultaneous wireless information and power transfer (SWIPT) is a valid mode in WET. Specifically, in SWIPT, the user divides the received RF signal into an information decoder (ID) and an energy harvester (EH) through power splitting (PS) or time switching (TS) \cite{7556958,7946258,7935500,8444683,8840901}. With MIMO technology, SWIPT can also be implemented through antenna switching or spatial switching. In antenna switching, each antenna element is switched dynamically between decoding/rectifying in the antenna domain \cite{6957150}. In spatial switching, information or energy is transmitted through eigenchannels obtained by eigenvalue decomposition of the MIMO channel matrix. According to the above-mentioned implementation technology of SWIPT, there have been many studies on the integration of SWIPT into existing communication technology \cite{8026127,9308990,9531372}. Power splitting factor and signal autocorrelation matrix are designed jointly to maximize the power harvested in the MIMO channel \cite{8026127}. Buckley \emph{et al.} proposed an energy receiving architecture under orthogonal frequency division multiplexing (OFDM) system \cite{9308990}. Under this architecture, the user performs energy harvesting and storage from the cyclic prefix of the signal. SWIPT in non-orthogonal multiple access (NOMA) network was studied in \cite{9531372}, authors consider energy harvested constraint and the quality-of-service (QoS) requirement of each user and minimize BS transmit power. For the various implementations of SWIPT mentioned above, there are also studies comparing these implementations in specific scenarios, especially PS and TS \cite{6893021,8630972,6760603}. The authors compare the attainable rate-energy trade-off in SWIPT-based communication systems for multiple-input single-output (MISO) channel \cite{6893021} and MIMO channel \cite{8630972}. Zhou \emph{et al.} considered the joint optimization of resource allocation and power splitting in the OFDM system \cite{6760603}. Jiang \emph{et al.} approximately obtained the optimal solution to the probability of information and energy coverage  for UAVs assisting SWIPT networks and verified it with the Monte Carlo method \cite{8880479}. All of the above works demonstrate that the performance of the PS scheme is better than that of the TS scheme. However, PS-based SWIPT can solve the energy shortage in IoT devices, but the energy consumption and cost of BSs also need to be considered urgently.
 
Considering the requirements to reduce the power consumption and cost of the BS, the recently proposed reconfigurable metasurface (RMS) may be a potential solution. RMS also known as reconfigurable intelligent surface (RIS), is an advanced technology that makes it possible to reconfigure wireless channels in wireless communications networks. RMS contains many passive elements with adjustable phase and amplitude. Since RMS is a passive communication equipment, it can only reflect or transmit signal and does not perform signal processing. RMS has the characteristics of low cost and easy deployment and is an environment-friendly communication device \cite{9119122,9716123,9676676}. Because of the above advantages of RMS, it has been widely studied in both academia and industry. Specifically, depending on the medium material, RMS is mainly divided into three types: reflective RMS \cite{9110912,9509394,9206080,9485102}, transmissive RMS \cite{9365009,9200683} and simultaneously transmitting and reflecting (STAR) RMS \cite{9863732,9740451}. For reflective RMS, it is also called intelligent reflecting surface (IRS) and used to improve the energy efficiency and spectral efficiency of communication networks, and RMS can make the system obtain obvious performance gains in the main communication scenarios. Zhang \emph{et al.} and Yang \emph{et al.} maximized the communication capacity of the IRS-assisted system in MIMO systems, respectively \cite{9110912}. Yang \emph{et al.} applied IRS to physical layer security to maximize the secret rate \cite{9206080}. For the transmissive RMS, it can solve the problem of blind coverage in the communication networks. Zeng \emph{et al.} evaluated the performance of the downlink RIS-assisted communication system and summarized the selection of the optimal working mode of RIS for a specific user location \cite{9365009}. Zhang \emph{et al.} proposed an intelligent omni-surface communication system, where transmissive elements adjust the phase of the received signal to improve network coverage \cite{9200683}. While STAR RMS can split the incident signal into transmitted and reflected signals, helping to achieve full spatial coverage on both sides of the surface. Wu \emph{et al.} studied the problem of resource allocation in STAR RMS-assisted multi-carrier communication networks \cite{9740451}. In the above researches, RMS is used as a communication auxiliary device for channel reconstruction and performance boost in two modes.

Furthermore, RMS can also be used as a transmitter, which is a very promising research direction. Tang \emph{et al.} implemented real-time communication of quadrature amplitude modulation (QAM)-MIMO by using reflective RMS and verified the theoretical model \cite{9133266}. In terms of transmitter design, transmissive RMS has better performance than reflective RMS, which is mainly because of the following two reasons \cite{bai2020high,wan2022space,liu2021multifunctional,7448838}. One of the reasons is that when RMS works in the reflective mode, the user and the feed antenna are located on the same side of the RMS, which makes the incident and the reflected electromagnetic (EM) waves to interfere with each other. Another reason is that the transmissive RMS transceiver can be designed with higher aperture efficiency and operating bandwidth \cite{bai2020high}. For the above reasons, applying transmissive RMS to multi-antenna transmitter designs is a potential technique in future wireless communications \cite{9570775}.

\newcounter{my5}
\begin{figure*}[!t]
	\normalsize
	\setcounter{my5}{\value{equation}}
	\setcounter{equation}{2}
	\begin{equation}\label{aod}
		\begin{aligned}
  & {{\mathbf{h}}_{k,\text{LoS}}}={{\left[ 1,{{e}^{-j\frac{2\pi }{\lambda }d\sin \theta _{k}^{\text{AoD}}\cos \phi _{k}^{\text{AoD}}}},\ldots ,{{e}^{-j\frac{2\pi }{\lambda }\left( {{N}_{x}}-1 \right)d\sin \theta _{k}^{\text{AoD}}\cos \phi _{k}^{\text{AoD}}}} \right]}^{T}} \!\otimes\! {{\left[ 1,{{e}^{-j\frac{2\pi }{\lambda }d\sin \theta _{k}^{\text{AoD}}\sin \phi _{k}^{\text{AoD}}}},\ldots ,{{e}^{-j\frac{2\pi }{\lambda }\left( {{N}_{z}}-1 \right)d\sin \theta _{k}^{\text{AoD}}\sin \phi _{k}^{\text{AoD}}}} \right]}^{T}}, \\ 
		\end{aligned}
	\end{equation}
	\setcounter{equation}{\value{my5}}
	\hrulefill
	\vspace*{4pt}
\end{figure*}

In view of the two important issues in IoT networks: the limited battery capacity of the devices and the excessive energy consumption of the BSs, we propose a downlink transmission design scheme for SWIPT networks based on the transmissive RMS transceiver. In order to make the design more practical, a nonlinear energy harvesting model is applied to this network model. Compared with the linear energy harvesting model, the nonlinear model has higher energy conversion efficiency \cite{7264986}. Considering the difficulty of channel estimation in RMS-assisted systems, the channel estimation error matrix is introduced into our model to simulate the impact of imperfect channel state information (CSI). In this paper, we aim to maximize the system sum-rate downlink by jointly optimizing RMS transmissive coefficient, power allocation, and power splitting ratio with the outage probability criterion. Given that the problem formulated is non-convex, it is necessary to design a reasonable and effective algorithm to solve it. The main contributions of this paper can be summarized as follows:

\begin{itemize}
\item We propose a novel transmissive RMS transceiver-enabled SWIPT network architecture, where the RMS is used as transceiver to implement beamforming. Specifically, RMS transmissive coefficient, transmit power allocation and power splitting ratio are designed jointly to maximize the system sum-rate. Taking into account the imperfect CSI, we use outage probability to measure QoS and energy harvested requirements, which can demonstrate the robustness of our design. However, it is non-trivial to directly obtain the global optimal solution to this problem since the high coupling of optimization variables.

\item We propose a joint optimization algorithm based on an alternating optimization (AO) framework to solve this formulated robust system sum-rate maximization problem. Specifically, the original problem is first transformed into a tractable problem. Then, the original problem is decoupled into three sub-problems with respect to transmit power allocation, power splitting ratio and RMS transmissive coefficient to be solved separately. Finally, we alternately optimize the three sub-problems till the entire problem converges.
 
\item Numerical results reveal the superior performance of the proposed algorithm in downlink multi-user SWIPT networks with transmissive RMS as transmitter. Specifically, the algorithm first has good convergence. Secondly, under the constraints of information and energy harvested requirements based on outage probability criterion, the robust joint optimization algorithm can improve the sum-rate of system compared to other benchmarks under the conditions of different number of RMS elements, number of users, and maximum transmit power.
\end{itemize}

The rest of this paper is as follows. In section II, we delineate the system model and optimization problem formulation in transmissive RMS transceiver-enabled SWIPT networks when considering the non-linear EH model and the imperfect CSI. Then, in section III, the proposed robust joint optimization algorithm is elaborated. Section IV reveals the performance superiority of the proposed algorithm compared to other benchmarks. Finally, section V concludes this paper.

\textit{Notations:} Matrices are represented by bold uppercase letters. Vectors are denoted by bold lowercase letters. Scalars are represented by standard lowercase letters. For a complex-valued scalar $x$, $\left| {x} \right|$ denotes its absolute value and for a complex-valued vector $\bf{x}$, $\left\| {\bf{x}} \right\|$ represents the Euclidean norm. For a general matrix $\bf{A}$, $\rm{rank(\bf{A})}$, ${\bf{A}}^H$, ${\bf{A}}_{m,n}$ and $\left\| {\bf{A}} \right\|$ denote its rank, conjugate transpose, $m,n$-th entry and matrix norm, respectively. For a square matrix $\bf{X}$, $\rm{tr(\bf{X})}$ and $\rm{rank(\bf{X})}$, denote its trace, rank, and ${\bf{X}} \succeq 0$ denotes that $\bf{X}$ is a positive semidefinite matrix. ${\mathbb{C}^{M \times N}}$ represents the ${M \times N}$ dimensional complex matrix space and $j$ is the imaginary unit. Finally, ${\cal C}{\cal N}\left( {\mu ,{\bf{C}}} \right)$ denotes the distribution of a circularly symmetric complex Gaussian (CSCG) random vector with mean $\mu$ and covariance matrix ${\bf{C}}$, and $ \sim $ stands for `distributed as'.

\section{System Model and Optimization Problem Formulation}
\subsection{System Model}
As shown in Fig. 1, the system model of transmissive RMS transceiver-enabled SWIPT networks is first introduced and it mainly includes a transmissive RMS transceiver and $K$ users with a single antenna. The transceiver is composed of a transmissive RMS with $N$ elements and a feed antenna. It is worth noting that although we are considering a transmissive RMS transceiver architecture, a portion of the electromagnetic waves emitted from the feed antenna will always be reflected. However, we can quantify this part of the reflected electromagnetic wave by a certain ratio, so it does not affect the algorithm design of this problem. In this paper, for the convenience of analysis, we assume that the electromagnetic wave is completely transmitted, i.e., no incident electromagnetic waves are reflected. The transmissive RMS is equipped with an intelligent controller which can control the amplitude and phase shift of all transmissive elements. We let $\mathbf{f}={{\left[ {{f}_{1}},...,{{f}_{N}} \right]}^{T}}\in {{\mathbb{C}}^{N\times 1}}$ represent the RMS transmissive coefficient vector at the transmitter, where ${{f}_{n}}={{\beta }_{n}}{{e}^{j{{\theta }_{n}}}}$ represents the amplitude and phase shift of the $n$-th element respectively, which should satisfy
\begin{figure}
	\centerline{\includegraphics[width=10cm]{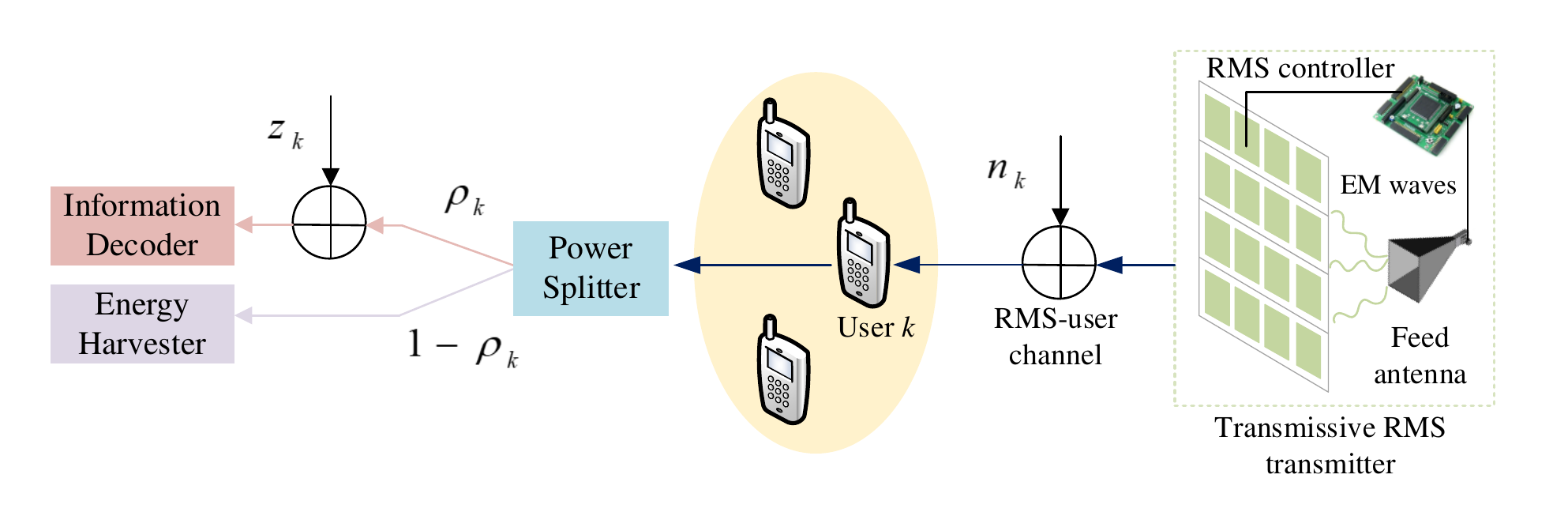}}
	\caption{Transmissive RMS transceiver-enabled SWIPT networks.}
	\label{Fig1}
\end{figure}
\begin{equation}
	{\left| {{f}_{n}} \right|}\le 1,\forall n.
\end{equation}
The channel from the RMS transceiver to the $k$-th user can be named as the RMS-user channel, and the channel gain can be denoted by $\mathbf{h}_{k}^{H}\in {{\mathbb{C}}^{1\times N}}$. For ease of analysis, all channels are assumed to be quasi-static flat fading, i.e., $\mathbf{h}_{k}^{H}$ is constant within each transmission time $T$. It is worth noting that the transmissive RMS transmits the signal passively and has no ability to actively send and receive signals. We assume that the communication works in time division duplex (TDD) mode, i.e., the channel estimation is completed in the uplink transmission. Downlink CSI can be obtained according to channel reciprocity. This paper assumes that the transmissive RMS transceiver cannot obtain the CSI perfectly, and the specific modeling is explained below. 

In this paper, we model the array of RMS as a uniform planar array (UPA), which is a more realistic array response, i.e., $N={{N}_{x}}\times {{N}_{z}}$, ${{N}_{x}}$ and ${{N}_{z}}$ denote the number of elements in the horizontal and vertical directions of the transmissive RMS, respectively. Herein, RMS-user channel is modeled as a Rice channel model, which can be given by
\setcounter{equation}{1}
\begin{equation}
{{\mathbf{h}}_{k}}=\sqrt{\beta {{\left( \frac{{{d}_{k}}}{{{d}_{0}}} \right)}^{-\alpha }}}\left( \sqrt{\frac{\kappa }{\kappa +1}}{{\mathbf{h}}_{k,\text{LoS}}}+\sqrt{\frac{1}{\kappa +1}}{{\mathbf{h}}_{k,\text{NLoS}}} \right),\forall k,
\end{equation}
where $\beta$ denotes the channel gain when the reference distance ${{d}_{0}}=1$ m, $\alpha$ is the path loss exponent between the RMS transceiver and the user, ${{d}_{k}}$ is the distance between RMS transceiver and the $k$-th user. $\kappa$ denotes the Rician
factor, ${{\mathbf{h}}_{k,\text{LoS}}}$ represents the LoS component, which can be determined by the Eq. (3) at the top of this page, where $\theta _{k}^{\rm{AoD}}$ and $\phi _{k}^{\rm{AoD}}$ are the vertical angle and horizontal angle of the angle-of-departure (AoD) at the RMS transceiver, respectively.
$d$ denotes the spacing between successive antenna elements and $\lambda$ denotes the carrier wavelength. ${{\mathbf{h}}_{k,\text{NLoS}}}$ represents the NLoS component and ${{\left[ {{\mathbf{h}}_{k,\text{NLoS}}} \right]}_{\left( {{n}_{x}}-1 \right){{N}_{z}}+{{n}_{z}}}}\sim \mathcal{C}\mathcal{N}\left( 0,1 \right)$ is
the $\left( {{n}_{x}}-1 \right){{N}_{z}}+{{n}_{z}}$ element of the vector ${{\mathbf{h}}_{k,\text{NLoS}}}$. Accordingly, the signal received by the $k$-th user can be denoted by
\setcounter{equation}{3}
\begin{equation}
	{{y}_{k}}=\mathbf{h}_{k}^{H}\mathbf{f}\sum\limits_{i=1}^{K}{\sqrt{{{p}_{i}}}{{s}_{i}}}+{{n}_{k}},\forall k,
\end{equation}
where ${{s}_{i}}$ denotes the signal from the RMS transceiver to the $i$-th user, Without loss of generality, we usually assume that it is an independent and identically distributed (i.i.d) CSCG random variable, i.e., ${{s}_{i}}\sim \mathcal{C}\mathcal{N}\left( 0,1 \right)$. ${{n}_{k}}$ represents additive white Gaussian noise (AWGN) introduced at the $k$-th user's receiving antenna, and it is also usually set assumed to be i.i.d CSCG variable, i.e., ${{n}_{k}}\sim \mathcal{C}\mathcal{N}\left( 0,\sigma _{k}^{2} \right)$. ${{p}_{k}}$ represents the power allocated to the $k$-th user and the following constraints should be satisfied
\begin{equation}
	{{p}_{k}}\ge 0,\forall k,
\end{equation}
and
\begin{equation}
	\sum\limits_{k=1}^{K}{{{p}_{k}}}\le {{P}_{\max }},
\end{equation}
where ${{P}_{\max }}$ is the maximum transmit power of transmissive RMS transceiver.

This paper considers transmissive RMS transceiver-enabled SWIPT networks. Specifically, from the received RF signal, each user adopts the PS protocol to coordinate energy harvesting and information decoding, i.e., each user's received signal is divided into the ID and EH by the power splitter. The $k$-th user divide the ${{\rho }_{k}}$ portion of the received signal power to ID and the rest $(1-{{\rho }_{k}})$ portion to EH. Therefore, the received signal for ID in the downlink of the $k$-th user is denoted by
\begin{equation}
	y_{k}^{\text{ID}}=\sqrt{{{\rho }_{k}}}{{y}_{k}}=\sqrt{{{\rho }_{k}}}\left( \mathbf{h}_{k}^{H}\mathbf{f}\sum\limits_{i=1}^{K}{\sqrt{{{p}_{i}}}{{s}_{i}}}+{{n}_{k}} \right)+{{z}_{k}},\forall k,
\end{equation}
where ${{z}_{k}}$ represents AWGN caused by the ID of the $k$-th user and it is set to be an i.i.d CSCG variable, ${{z}_{k}}\sim \mathcal{C}\mathcal{N}\left( 0,\delta _{k}^{2} \right)$. Then, the signal to interference plus noise ratio (SINR) of the $k$-th user is denoted by
\begin{equation}
	\text{SIN}{{\text{R}}_{k}}=\frac{{{\rho }_{k}}{{p}_{k}}{{\left| \mathbf{h}_{k}^{H}\mathbf{f} \right|}^{2}}}{{{\rho }_{k}}\sum\limits_{i\ne k}{{{p}_{i}}{{\left| \mathbf{h}_{k}^{H}\mathbf{f} \right|}^{2}}+{{\rho }_{k}}{{\sigma }^{2}}+\delta _{k}^{2}}},\forall k.
\end{equation}
In addition, for the $k$-th user, the received signal for EH in the downlink can be given by
\begin{equation}
	y_{k}^{\text{EH}}=\sqrt{1-{{\rho }_{k}}}{{y}_{k}}=\sqrt{1-{{\rho }_{k}}}\left( \mathbf{h}_{k}^{H}\mathbf{f}\sum\limits_{i=1}^{K}{\sqrt{{{p}_{i}}}{{s}_{i}}}+{{n}_{k}} \right),\forall k.
\end{equation}
Accordingly, the power obtained by the $k$-th user for EH is given by
\begin{equation}
	p_{k}^{\text{EH}}=\mathbb{E}\left\{ {{\left| y_{k}^{\text{EH}} \right|}^{2}} \right\}=\left( 1-{{\rho }_{k}} \right)\left( \sum\limits_{i=1}^{K}{{{p}_{i}}{{\left| \mathbf{h}_{k}^{H}\mathbf{f} \right|}^{2}}}+\sigma _{k}^{2} \right),\forall k.
\end{equation}
In this paper, a more practical non-linear energy harvested model is adopted. Hence, the power harvested by the $k$-th user can be expressed as
\begin{equation}
	\Psi \left( p_{k}^{\text{EH}} \right)=\left( \frac{{{\partial }_{k}}}{{{X}_{k}}\left( 1+\exp \left( -{{a}_{k}}\left( p_{k}^{\text{EH}}-{{b}_{k}} \right) \right) \right)}-{{Y}_{k}} \right),\forall k,
\end{equation}
where ${{\partial }_{k}}$ represents the maximum energy harvested of the $k$-th user, ${{a}_{k}}$ and ${{b}_{k}}$ are specific parameters related to the circuit. ${{X}_{k}}={\exp \left( {{a}_{k}}{{b}_{k}} \right)}/{\left( 1+\exp \left( {{a}_{k}}{{b}_{k}} \right) \right)}$ and ${{Y}_{k}}={{{\partial }_{k}}}/{\exp \left( {{a}_{k}}{{b}_{k}} \right)}$.
We consider that under normalized time, the energy harvested by the $k$-th user can be given by
\begin{equation}
	{{E}_{k}}=\Psi \left( p_{k}^{\text{EH}} \right),\forall k.
\end{equation}
Let ${{\mathbf{\Phi }}_{k}}=\mathbb{E}\left\{ {{\mathbf{h}}_{k}}\mathbf{h}_{k}^{H} \right\}={{\mathbf{h}}_{k}}\mathbf{h}_{k}^{H}\in {{\mathbb{C}}^{N\times N}}$ represent the channel covariance matrix of the $k$-th user in the downlink\footnote{We consider a quasi-static channel model, i.e., during each transmission time duration $T$, ${\mathbf{h}}_{k}$ is a constant. Therefore, we use the instantaneous value of the channel gain to compute the channel covariance matrix instead of the expectation operator.}. Then, the SINR of the $k$-th user can be expressed by the channel covariance matrix as
\begin{equation}
	\text{SIN}{{\text{R}}_{k}}=\frac{{{\rho }_{k}}{{p}_{k}}{\rm{tr}}\left( {{\mathbf{\Phi }}_{k}}\mathbf{F} \right)}{{{\rho }_{k}}\sum\limits_{i\ne k}{{{p}_{i}}{\rm{tr}}\left( {{\mathbf{\Phi }}_{k}}\mathbf{F} \right)+{{\rho }_{k}}\sigma _{k}^{2}+\delta _{k}^{2}}},\forall k,
\end{equation}
where $\mathbf{F}=\mathbf{f}{{\mathbf{f}}^{H}}\in {{\mathbb{C}}^{N\times N}}$ and it should satisfy  $\rm{rank}\left( \mathbf{F} \right)=1,\mathbf{F}\succeq 0$ and ${{\mathbf{F}}_{n,n}}\le 1,\forall n$. 
In addition, the energy harvested of the $k$-th user is further denoted by
\begin{equation}
	{{E}_{k}}=\Psi \left( \left( 1-{{\rho }_{k}} \right)\left( \sum\limits_{i=1}^{K}{{{p}_{i}}{\rm{tr}}\left( {{\mathbf{\Phi }}_{k}}\mathbf{F} \right)}+\sigma _{k}^{2} \right) \right),\forall k.
\end{equation}

To make the model more realistic, we consider that the CSI of the downlink cannot be obtained accurately, i.e., in the case of imperfect CSI. Specifically,  the channel covariance matrix is assumed to be expressed as ${{\mathbf{\Phi }}_{k}}+\Delta {{\mathbf{\Phi }}_{k}}$, where ${{\mathbf{\Phi }}_{k}}\in {{\mathbb{C}}^{N\times N}}$ denotes the covariance matrix of the estimated channel in the downlink and $\Delta {{\mathbf{\Phi }}_{k}}\in {{\mathbb{C}}^{N\times N}}$ is the error matrix corresponding to the estimated error of ${{\mathbf{\Phi }}_{k}}$, which can also be called the uncertainty matrix, because it represents the difference between the estimated value and the true value \cite{4350296}. Note that ${{\mathbf{\Phi }}_{k}}$ and $\Delta {{\mathbf{\Phi }}_{k}}$ are Hermitian matrices, then the SINR and the energy harvested for the $k$-th user is denoted by
\begin{equation}
	\text{SIN}{{\text{R}}_{k}}=\frac{{{\rho }_{k}}{{p}_{k}}{\rm{tr}}\left( \left( {{\mathbf{\Phi }}_{k}}+\Delta {{\mathbf{\Phi } }_{k}} \right)\mathbf{F} \right)}{{{\rho }_{k}}\sum\limits_{i\ne k}{{{p}_{i}}}{\rm{tr}}\left( \left( {\mathbf{\Phi }_{k}}+\Delta {{\mathbf{\mathbf{\Phi } } }_{k}} \right)\mathbf{F} \right)+{{\rho }_{k}}\sigma _{k}^{2}+\delta _{k}^{2}},\forall k,	
\end{equation}
and
\begin{equation}
	{{E}_{k}}=\Psi \left( \left( 1-{{\rho }_{k}} \right)\left( \sum\limits_{i=1}^{K}{{{p}_{i}}\text{tr}\left( \left( {{\mathbf{\Phi }}_{k}}+\Delta {{\mathbf{\Phi }}_{k}} \right)\mathbf{F} \right)}+\sigma _{k}^{2} \right) \right),\forall k.
\end{equation}
Accordingly, the $k$-th user's achievable rate (bps/Hz) is expressed as
\begin{equation}
	{{R}_{k}}={{\log }_{2}}\left( 1+\text{SIN}{{\text{R}}_{k}} \right),\forall k.
\end{equation}

Since random matrix variable terms are involved in $R_k$, we take its expectation, which can be defined as $\mathbb{E}\left\{ {{R}_{k}} \right\}$. However, we can't use general methods to directly obtain a closed-form expression for the expectation. To solve this problem, we approximate the expectation of the achievable rate by applying $\textbf{Proposition 1}$ below.
\newcounter{my6}
\begin{figure*}[!t]
	\normalsize
	\setcounter{my6}{\value{equation}}
	\setcounter{equation}{19}
	\begin{equation}\label{aod}
		\mathcal{O}_{k}^{\text{ID}}=\Pr \left\{ \text{SINR}_{k}\le {{\gamma }_{th}} \right\}=\Pr \left\{ \frac{{{\rho }_{k}}{{p}_{k}}{\rm{tr}}\left( \left( {{\mathbf{\Phi }}_{k}}+\Delta {{\mathbf{\Phi }}_{k}} \right)\mathbf{F} \right)}{{{\rho }_{k}}\sum\limits_{i\ne k}{{{p}_{i}}{\rm{tr}}\left( \left( {{\mathbf{\Phi }}_{k}}+\Delta {{\mathbf{\Phi }}_{k}} \right)\mathbf{F} \right)+{{\rho }_{k}}\sigma _{k}^{2}+\delta _{k}^{2}}}\le {{\gamma }_{th}} \right\},\forall k,
	\end{equation}
	\setcounter{equation}{\value{my6}}
	\hrulefill
	\vspace*{4pt}
\end{figure*}
\newcounter{my7}
\begin{figure*}[!t]
	\normalsize
	\setcounter{my7}{\value{equation}}
	\setcounter{equation}{20}
	\begin{equation}\label{aod}
	\mathcal{O}_{k}^{\text{EH}}=\text{Pr}\left\{ {{E}_{k}}\le {{E}_{th}} \right\}=\Pr \left\{ \Psi \left( \left( 1-{{\rho }_{k}} \right)\left( \sum\limits_{i=1}^{K}{{{p}_{i}}{\rm{tr}}\left( \left( {{\mathbf{\Phi }}_{k}}+\Delta {{\mathbf{\Phi }}_{k}} \right)\mathbf{F} \right)}+\sigma _{k}^{2} \right) \right)\le {{E}_{th}} \right\},\forall k.
	\end{equation}
	\setcounter{equation}{\value{my7}}
	\hrulefill
	\vspace*{4pt}
\end{figure*}

$\textbf{Proposition 1:}$ For any $a$ and $b$, if $X$ is a random variable term or contains a random variable term, the following approximation holds,

\begin{equation}
\mathbb{E}\left\{ {{\log }_{2}}\left( 1+\frac{aX}{bX+1} \right) \right\}\approx \mathbb{E}\left\{ {{\log }_{2}}\left( 1+\frac{\mathbb{E}\left\{ aX \right\}}{\mathbb{E}\left\{ bX \right\}+1} \right) \right\}.
\end{equation}

\emph{Proof}: The proof of this formula is similar to the proof of \textbf{Theorem 1} in Ref. \cite{9121338} and here the proof is omitted. $\hfill\blacksquare$

For the convenience of analysis, we assume that $\Delta {{\mathbf{\Phi }}_{k}}$ is a Hermitian matrix, and the elements on the diagonal are i.i.d. cyclic symmetric real Gaussian random variables with zero mean and $\sigma _{\Phi }^{2}$ variance. Other elements are i.i.d. CSCG random variables with zero mean and $\sigma _{\Phi }^{2}$ variance. According to the $\textbf{Proposition 1}$, we can take that the expectation of the $k$-th user's achievable rate as follows
\begin{equation}
	\mathbb{E}\left\{ {{R}_{k}} \right\}\approx {{\log }_{2}}\left( 1+\frac{{{\rho }_{k}}{{p}_{k}}{\rm{tr}}\left( {{\mathbf{\Phi }}_{k}}\mathbf{F} \right)}{{{\rho }_{k}}\sum\limits_{i\ne k}{{{p}_{i}}{\rm{tr}}\left( {{\mathbf{\Phi }}_{k}}\mathbf{F} \right)+{{\rho }_{k}}\sigma _{k}^{2}+\delta _{k}^{2}}} \right),\forall k.
\end{equation}

Considering imperfect CSI, the user's SINR is a random variable, 
which means that we can only express the information and energy harvested requirement with outage probability. We define the information outage probability of the $k$-th user as the probability that its SINR is smaller than the threshold ${{\gamma }_{th}}$, which can be expressed as the Eq. (20), where $\Pr \left\{ \cdot  \right\}$ is the probability operator. Similarly, energy harvested outage probability is defined as the probability that the energy harvested is lower than the threshold ${{E}_{th}}$, which can be expressed as the Eq. (21).

\subsection{Problem Formulation}
Let $\bm{\rho }=\left[ {{{{\rho}} }_{1}},...,{{\rho }_{K}} \right]$, $\mathbf{p}=\left[ {{p}_{1}},...,{{p}_{K}} \right]$. We consider that the information outage probability of each user is not greater than  ${{\zeta }_{k}}$, and the energy outage probability of each user is not greater than ${{\varepsilon }_{k}}$. By jointly optimizing the power splitting ratio $\bm{\rho }$, RMS transmissive coefficient $\mathbf{F}$ and the transmit power allocation $\mathbf{p}$, the expectation of the system sum-rate is maximized. Therefore, the original problem P0 can be expressed as
\setcounter{equation}{21}
\begin{subequations}\label{p0}
	\begin{align}
		\text{P0}:\qquad&\underset{\mathbf{{{\bm{\rho}}} },\mathbf{p},\mathbf{F}}{\mathop{\max }}\,\text{ }\sum\limits_{k=1}^{K}{{{\log }_{2}}\left( 1+\frac{{{\rho }_{k}}{{p}_{k}}{\rm{tr}}\left( {{\mathbf{\Phi }}_{k}}\mathbf{F} \right)}{{{\rho }_{k}}\sum\limits_{i\ne k}{{{p}_{i}}{\rm{tr}}\left( {{\mathbf{\Phi }}_{k}}\mathbf{F} \right)+{{\rho }_{k}}\sigma _{k}^{2}+\delta _{k}^{2}}} \right)}, \nonumber\\ 
		\rm{s.t.}\qquad&{{p}_{k}}\ge 0,\forall k, \\ 
		&\sum\limits_{k=1}^{K}{{{p}_{k}}}\le {{P}_{\max }}, \\ 
		& 0\le {{\rho }_{k}}\le 1,\forall k, \\ 
		& \text{Pr}\left\{ \text{SIN}{{\text{R}}_{k}}\le {{\gamma }_{th}} \right\}\le {{\zeta }_{k}},\forall k, \\ 
		& \text{Pr}\left\{ {{E}_{k}}\le {{E}_{th}} \right\}\le {{\varepsilon }_{k}},\forall k, \\ 
		& {{\mathbf{F}}_{n,n}}\le 1,\forall n, \\ 
		& \mathbf{F}\succeq 0, \\ 
		& \rm{rank}\left( \mathbf{F} \right)=1, 
	\end{align}
\end{subequations}
where constraint (\ref{p0}a) and constraint (\ref{p0}b) are the transmit power allocation constraints of transmissive RMS transceiver, constraint (\ref{p0}c) is the power splitting ratio constraint of each user. To guarantee the QoS of user information and energy harvesting at the same time, constraint (\ref{p0}d) ensures that the information outage probability of each user is not greater than ${{\zeta }_{k}}$ and constraint (\ref{p0}e) ensures that the energy harvesting outage probability of each user is not greater than ${{\varepsilon }_{k}}$. Constraints (\ref{p0}f)-(\ref{p0}h) are RMS transmissive coefficient constraints.

As can be observed, the original problem P0 is non-convex for following several reasons: First, the highly coupled variables make the objective function non-concave. In addition, constraints (\ref{p0}d) and (\ref{p0}e) are constraints based on the outage probability criterion, which are difficult to handle directly. Finally, a non-convex rank-one constraint (\ref{p0}h) is introduced after the RMS transmissive coefficient vector is lifted to a matrix. Therefore, solving this problem is challenging.
\newcounter{my8}
\begin{figure*}[!t]
	\normalsize
	\setcounter{my8}{\value{equation}}
	\setcounter{equation}{22}
	\begin{equation}\label{aod}
		\mathcal{O}_{k}^{\text{ID}}=\Pr \left\{ {{\rho }_{k}}{{p}_{k}}{\rm{tr}}\left( \left( {{\mathbf{\Phi }}_{k}}+\Delta {{\mathbf{\Phi }}_{k}} \right)\mathbf{F} \right)\le {{\rho }_{k}}{{\gamma }_{th}}\sum\limits_{i\ne k}{{{p}_{i}}{\rm{tr}}\left( \left( {{\mathbf{\Phi }}_{k}}+\Delta {{\mathbf{\Phi }}_{k}} \right)\mathbf{F} \right)+{{\rho }_{k}}{{\gamma }_{th}}\sigma _{k}^{2}+{{\gamma }_{th}}\delta _{k}^{2}} \right\},\forall k. 
	\end{equation}
	\setcounter{equation}{\value{my8}}
	\hrulefill
	\vspace*{4pt}
\end{figure*}
\section{Robust Joint Optimization Algorithm Design in Transmissive RMS Transceiver-Enabled SWIPT Networks}
\subsection{Problem Transformation}
Obviously, the problem P0 is a non-convex optimization problem and needs to be transformed into a tractable convex problem. Next, we reformulate the probability constraint (21d) through a statistical model. Herein, $\mathcal{O}_{k}^{\text{ID}}$ can be rewritten as Eq. (23) on the top of the next page. We introduce the auxiliary matrix
\setcounter{equation}{23}
\begin{equation}
	{{\mathbf{\tilde{\Phi }}}_{k}}={{\rho }_{k}}{{p}_{k}}{{\mathbf{\Phi }}_{k}}-{{\rho }_{k}}{{\gamma }_{th}}\sum\limits_{i\ne k}{{{p}_{i}}{{\mathbf{\Phi }}_{k}}},\forall k,
\end{equation}
and
\begin{equation}
	\Delta {{\mathbf{\tilde{\Phi }}}_{k}}={{\rho }_{k}}{{p}_{k}}\Delta {{\mathbf{\Phi }}_{k}}-{{\rho }_{k}}{{\gamma }_{th}}\sum\limits_{i\ne k}{{{p}_{i}}\Delta {{\mathbf{\Phi }}_{k}}},\forall k.
\end{equation}
Then the information outage probability of the $k$-th user can be given by
\begin{equation}
	\mathcal{O}_{k}^{\text{ID}}=\Pr \left\{ {\rm{tr}}\left( \left( {{{\mathbf{\tilde{\Phi }}}}_{k}}+\Delta {{{\mathbf{\tilde{\Phi }}}}_{k}} \right)\mathbf{F} \right)\le {{\rho }_{k}}{{\gamma }_{th}}\sigma _{k}^{2}+{{\gamma }_{th}}\delta _{k}^{2} \right\},\forall k. 
\end{equation}
We define a random variable ${{\chi }_{k}}={\rm{tr}}\left( \left( {{{\mathbf{\tilde{\Phi }}}}_{k}}+\Delta {{{\mathbf{\tilde{\Phi }}}}_{k}} \right)\mathbf{F} \right),\forall k$ and an intermediate variable to be optimized ${{c}_{k}}={{\rho }_{k}}{{\gamma }_{th}}\sigma _{k}^{2}+{{\gamma }_{th}}\delta _{k}^{2},\forall k$. Since ${{\mathbf{\tilde{\Phi }}}_{k}}$, $\Delta {{\mathbf{\tilde{\Phi }}}_{k}}$  and $\mathbf{F}$ are all Hermitian matrices, the following $\textbf{Proposition 2}$ can be cited for the probability distribution analysis of ${{\chi }_{k}}$.
\newcounter{my9}
\begin{figure*}[!t]
	\normalsize
	\setcounter{my9}{\value{equation}}
	\setcounter{equation}{29}
	\begin{equation}\label{aod}
		\mathcal{O}_{k}^{\text{ID}}=\Pr \left\{ {{\chi }_{k}}\le {{c}_{k}} \right\}=\int_{-\infty }^{{{c}_{k}}}{\frac{1}{\sqrt{2\pi }{{\sigma }_{e,k}}\left\| \mathbf{F} \right\|}}\exp \left( -\frac{{{\left( {{\chi }_{k}}-{\rm{tr}}\left( {{{\mathbf{\tilde{\Phi }}}}_{k}}\mathbf{F} \right) \right)}^{2}}}{2\sigma _{e,k}^{2}{{\left\| \mathbf{F} \right\|}^{2}}} \right)d{{\chi }_{k}},\forall k,
	\end{equation}
	\setcounter{equation}{\value{my9}}
	\hrulefill
	\vspace*{4pt}
\end{figure*}
\newcounter{my10}
\begin{figure*}[!t]
	\normalsize
	\setcounter{my10}{\value{equation}}
	\setcounter{equation}{34}
	\begin{equation}\label{aod}
		\mathcal{O}_{k}^{\text{EH}}	=\Pr \left\{ \Psi \left( \left( 1-{{\rho }_{k}} \right)\left( \sum\limits_{i=1}^{K}{{{p}_{i}}{\rm{tr}}\left( \left( {{\mathbf{\Phi }}_{k}}+\Delta {{\mathbf{\Phi }}_{k}} \right)\mathbf{F} \right)}+\sigma _{e,k}^{2} \right) \right)\le {{E}_{th}} \right\} =\Pr \left\{{\rm{tr}}\left( \left( {{{\mathbf{\overset{\scriptscriptstyle\frown}{\Phi }}}}_{k}}+\Delta {{{\mathbf{\overset{\scriptscriptstyle\frown}{\Phi }}}}_{k}} \right)\mathbf{F} \right)\le {{\varphi }_{k}} \right\}, \forall k,
	\end{equation}
	\setcounter{equation}{\value{my10}}
	\hrulefill
	\vspace*{4pt}
\end{figure*}

\textbf{Proposition 2:} if $\mathbf{X}$ is a random matrix with CSCG random elements with 0 mean and variance $\sigma _{x}^{2}$, for any deterministic matrix $\mathbf{Y}$, the following formula is established
\begin{equation}
	\rm{tr}\left( \mathbf{YX} \right)\sim \mathcal{C}\mathcal{N}\left( \text{0,}\sigma _{x}^{2}\text{tr}\left( \mathbf{Y}{{\mathbf{Y}}^{H}} \right) \right).
\end{equation}
According to \textbf{Proposition 2} \cite{1379113}, we can obtain 
${{\chi}_{k}}\sim\mathcal{C}\mathcal{N}\left( \text{tr}\left({{{\tilde{\Phi }}}_{k}}\mathbf{F} \right),\sigma _{e,k}^{2}\rm{tr}\left( \mathbf{F}{{\mathbf{F}}^{\emph{H}}} \right) \right)$, where $\sigma _{e,k}^{2}$ can be given by
\begin{equation}
	\sigma _{e,k}^{2}=\rho _{k}^{2}p_{k}^{2}\sigma _{\Phi }^{2}+\rho _{k}^{2}\gamma _{th}^{2}\sum\limits_{i\ne k}{p_{i}^{2}}\sigma _{\Phi }^{2}, \forall k,
\end{equation}
then,
\begin{equation}
	\sigma _{e,k}^{2}=\rho _{k}^{2}\sigma _{\Phi }^{2}\left( p_{k}^{2}+\gamma _{th}^{2}\sum\limits_{i\ne k}{p_{i}^{2}} \right), \forall k.
\end{equation}
Therefore, the information outage probability of the $k$-th user ${{\mathcal{O}}_{k}^{\text{ID}}}$ can be obtained by the Eq. (30), where ${{\left\| \mathbf{F} \right\|}^{2}}=\rm{tr}\left( \mathbf{F}{{\mathbf{F}}^{{\emph{H}}}} \right)$.
According to the definition of the error function
\setcounter{equation}{30}
\begin{equation}
	erf\left( x \right)=\frac{2}{\sqrt{\pi }}\int_{0}^{x}{\exp \left( -{{u}^{2}} \right)}du,
\end{equation}
the information outage probability of the $k$-th user can finally be given by
\begin{equation}
	\mathcal{O}_{k}^{\text{ID}}=\frac{1}{2}-\frac{1}{2}erf\left( \frac{{\rm{tr}}\left( {{{\mathbf{\tilde{\Phi }}}}_{k}}\mathbf{F} \right)-{{c}_{k}}}{\sqrt{2}{{\sigma }_{e,k}}\left\| \mathbf{F} \right\|} \right),\forall k.
\end{equation}
Thus, constraint (\ref{p0}d) can be rewritten as
\begin{equation}
	\frac{1}{2}-\frac{1}{2}erf\left( \frac{{\rm{tr}}\left( {{{\mathbf{\tilde{\Phi }}}}_{k}}\mathbf{F} \right)-{{c}_{k}}}{\sqrt{2}{{\sigma }_{e,k}}\left\| \mathbf{F} \right\|} \right)\le {{\zeta }_{k}}, \forall k.\
\end{equation}
This formula can be converted to
\begin{equation}
	{\rm{tr}}\left( {{{\mathbf{\tilde{\Phi }}}}_{k}}\mathbf{F} \right)-{{c}_{k}}\ge \sqrt{2}{{\sigma }_{e,k}}\left\| \mathbf{F} \right\|er{{f}^{-1}}\left( 1-2{{\zeta }_{k}} \right), \forall k.	
\end{equation}
\newcounter{my11}
\begin{figure*}[!t]
	\normalsize
	\setcounter{my11}{\value{equation}}
	\setcounter{equation}{36}
	\begin{equation}\label{aod}
		\mathcal{O}_{k}^{\text{EH}}=\Pr \left\{ {{\varpi }_{k}}\le {{\varphi }_{k}} \right\}=\int_{-\infty }^{{{\varphi }_{k}}}{\frac{1}{\sqrt{2\pi }{{\beta }_{e,k}}\left\| \mathbf{F} \right\|}}\exp \left( -\frac{{{\left( {{\varpi }_{k}}-\text{tr}\left( {{{\mathbf{\overset{\scriptscriptstyle\frown}{\Phi }}}}_{k}}\mathbf{F} \right) \right)}^{2}}}{2\beta _{e,k}^{2}{{\left\| \mathbf{F} \right\|}^{2}}} \right)d{{\varpi }_{k}} =\frac{1}{2}-\frac{1}{2}erf\left( \frac{{\rm{tr}}\left( {{{\mathbf{\overset{\scriptscriptstyle\frown}{\Phi }}}}_{k}}\mathbf{F} \right)-{{\varphi }_{k}}}{\sqrt{2}{{\beta}_{e,k}}\left\| \mathbf{F} \right\|} \right), \forall k.
	\end{equation}
	\setcounter{equation}{\value{my11}}
	\hrulefill
	\vspace*{4pt}
\end{figure*}
\newcounter{my12}
\begin{figure*}[!t]
	\normalsize
	\setcounter{my12}{\value{equation}}
	\setcounter{equation}{40}
	\begin{equation}\label{aod}
      \sum\limits_{k = 1}^K {\left( {{{\log }_2}\left( {\sum\limits_{i = 1}^K {{\rho _k}{p_i}{\rm{tr}}\left( {{{\bf{\Phi }}_k}{\bf{F}}} \right) + {\rho _k}\sigma _k^2 + \delta _k^2} } \right) - {{\log }_2}\left( {{\rho _k}\sum\limits_{i \ne k} {{p_i}{\rm{tr}}\left( {{{\bf{\Phi }}_k}{\bf{F}}} \right) + {\rho _k}\sigma _k^2 + \delta _k^2} } \right)} \right)} = \sum\limits_{k = 1}^K {\left( {{g_k}\left( {\bf{F}} \right) - {{\bar g}_k}\left( {\bf{F}} \right)} \right),\forall k.} 
	\end{equation}
	\setcounter{equation}{\value{my12}}
	\hrulefill
	\vspace*{4pt}
\end{figure*}
Similarly, the $k$-th user's energy outage probability $\mathcal{O}_{k}^{\text{EH}}$ is denoted by Eq. (35), where ${{\mathbf{\overset{\scriptscriptstyle\frown}{\Phi }}}_{k}}=\left( 1-{{\rho }_{k}} \right)\sum\limits_{i=1}^{K}{{{p}_{i}}{{\mathbf{\Phi }}_{k}}}$, $\Delta {{\mathbf{\overset{\scriptscriptstyle\frown}{\Phi }}}_{k}}=\left( 1-{{\rho }_{k}} \right)\sum\limits_{i=1}^{K}{{{p}_{i}}\Delta {{\mathbf{\Phi }}_{k}}}$ and ${{\varphi }_{k}}={{\Psi }^{-1}}\left( {{E}_{th}} \right)-\left( 1-{{\rho }_{k}} \right)\sigma _{k}^{2}$.
Then, we define a random variable ${{\varpi }_{k}}={\rm{tr}}\left( \left( {{{\mathbf{\overset{\scriptscriptstyle\frown}{\Phi }}}}_{k}}+\Delta {{{\mathbf{\overset{\scriptscriptstyle\frown}{\Phi }}}}_{k}} \right)\mathbf{F} \right)$. According to \textbf{Proposition 2}, we can obtain ${{\varpi }_{k}}\sim \mathcal{C}\mathcal{N}\left( {\rm{tr}}\left( {{{\mathbf{\overset{\scriptscriptstyle\frown}{\Phi }}}}_{k}}\mathbf{F} \right),\beta _{e,k}^{2}{\rm{tr}}\left( \mathbf{F}{{\mathbf{F}}^{H}} \right) \right)$, where $\beta _{e,k}^{2}$ can be given by
\setcounter{equation}{35}
\begin{equation}
	\beta _{e,k}^{2}={{\left( 1-{{\rho }_{k}} \right)}^{2}}\sum\limits_{i=1}^{K}{{{p}_{i}}^{2}\sigma _{\Phi }^{2}}, \forall k.
\end{equation}
Therefore, the $k$-th user's energy outage probability $\mathcal{O}_{k}^{\text{EH}}$ is obtained by Eq. (37) on the top of the next page. Thus, the constraint (\ref{p0}e) can be rewritten as
\setcounter{equation}{37}
\begin{equation}\label{aa}
	\frac{1}{2}-\frac{1}{2}erf\left( \frac{{\rm{tr}}\left( {{{\mathbf{\overset{\scriptscriptstyle\frown}{\Phi }}}}_{k}}\mathbf{F} \right)-{{\varphi }_{k}}}{\sqrt{2}{{\beta }_{e,k}}\left\| \mathbf{F} \right\|} \right)\le {{\varepsilon }_{k}}, \forall k.
\end{equation}
This formula can be converted to
\begin{equation}
	{\rm{tr}}\left( {{{\mathbf{\overset{\scriptscriptstyle\frown}{\Phi }}}}_{k}}\mathbf{F} \right)-{{\varphi }_{k}}\ge \sqrt{2}{{\beta }_{e,k}}\left\| \mathbf{F} \right\|er{{f}^{-1}}\left( 1-2{{\varepsilon }_{k}} \right), \forall k.
\end{equation}
Hence, we can transform the problem P0 into problem P1, which can be given by
\begin{subequations}
	\begin{align}
		\text{P}1:\qquad&\underset{\mathbf{{\bm{\rho}} },\mathbf{p},\mathbf{F}}{\mathop{\max }}\,\text{ }\sum\limits_{k=1}^{K}{{{\log }_{2}}\left( 1+\frac{{{\rho }_{k}}{{p}_{k}}{\rm{tr}}\left( {{\mathbf{\Phi }}_{k}}\mathbf{F} \right)}{{{\rho }_{k}}\sum\limits_{i\ne k}{{{p}_{i}}{\rm{tr}}\left( {{\mathbf{\Phi }}_{k}}\mathbf{F} \right)+{{\rho }_{k}}\sigma _{k}^{2}+\delta _{k}^{2}}} \right)}, \nonumber\\ 
		\rm{s.t.}\qquad&{{p}_{k}}\ge 0,\forall k, \\ 
		& \sum\limits_{k=1}^{K}{{{p}_{k}}}\le {{P}_{\max }}, \\ 
		& 0\le {{\rho }_{k}}\le 1,\forall k, \\ 
		& {\rm{tr}}\left( {{{\mathbf{\tilde{\Phi }}}}_{k}}\mathbf{F} \right)-{{c}_{k}}\ge \sqrt{2}{{\sigma }_{e,k}}\left\| \mathbf{F} \right\|er{{f}^{-1}}\left( 1-2{{\zeta }_{k}} \right),\forall k, \\ 
		& {\rm{tr}}\left( {{{\mathbf{\overset{\scriptscriptstyle\frown}{\Phi }}}}_{k}}\mathbf{F} \right)-{{\varphi }_{k}}\ge \sqrt{2}{{\beta }_{e,k}}\left\| \mathbf{F} \right\|er{{f}^{-1}}\left( 1-2{{\varepsilon }_{k}} \right),\forall k, \\ 
		& {{\mathbf{F}}_{n,n}}\le 1,\forall n, \\ 
		& \mathbf{F}\succeq 0, \\ 
		& \rm{rank}\left( \mathbf{F} \right)=1.  
	\end{align}
\end{subequations}

After the original problem is transformed, the AO framework can be implemented to decouple the problem P1 into three sub-problems: RMS transmissive coefficient optimization, transmit power allocation optimization, and power splitting ratio optimization. Then three non-convex sub-problems are transformed into convex sub-problems by applying DC programming and SCA, respectively. Next, by alternately optimizing these three sub-problems to reach convergence, the final RMS transmissive coefficient, transmit power allocation, and power splitting ratio scheme can be obtained.

\subsection{RMS Transmissive Coefficient Optimization}
In this subsection, we first fix the power splitting ratio $\bm{\rho }$ and transmit power allocation $\mathbf{p}$, and optimize the RMS transmissive coefficient $\mathbf{F}$. The objective function can be expressed as the Eq. (41), which is the difference of two concave functions with respect to (w.r.t) $\bf{F}$, which are not concave. Herein, we approximate ${{\bar{g}}_{k}}\left( \mathbf{F} \right)$ linearly by SCA as follows
\setcounter{equation}{41}
\begin{equation}
	\begin{aligned}
		{{\bar{g}}_{k}}\left( \mathbf{F} \right)&\le {{\bar{g}}_{k}}\left( {{\mathbf{F}}^{r}} \right)+{\rm{tr}}\left( {{\left( {{\nabla }_{\mathbf{F}}}{{{\bar{g}}}_{k}}\left( {{\mathbf{F}}^{r}} \right) \right)}^{H}}\left( \mathbf{F}-{{\mathbf{F}}^{r}} \right) \right)\\
		&\triangleq {{\bar{g}}_{k}}{{\left( \mathbf{F} \right)}^{ub}},\forall k,
	\end{aligned}
\end{equation}
with
\begin{equation}
	{{\nabla }_{\mathbf{F}}}{{\bar{g}}_{k}}({{\mathbf{F}}^{r}})=\frac{{{\rho }_{k}}\sum\limits_{i\ne k}{{{p}_{i}}\mathbf{\Phi }_{k}^{H}}}{\left( {{\rho }_{k}}\sum\limits_{i\ne k}{{{p}_{i}}\rm{tr}\left( {{\mathbf{\Phi }}_{k}}\mathbf{F}^{r} \right)+{{\rho }_{k}}\sigma _{k}^{2}+\delta _{k}^{2}} \right)\ln 2},\forall k,
\end{equation}
where ${{\mathbf{F}}^{r}}$ represents the value at the $r$-th SCA iteration. Therefore, the problem P1 can be approximately expressed as follows
\begin{subequations}\label{p2}
	\begin{align}
		\text{P2: }\qquad &\underset{\mathbf{F}}{\mathop{\text{max}}}\,\text{  }\sum\limits_{k=1}^{K}{\left( {{g}_{k}}\left( \mathbf{F} \right)-{{{\bar{g}}}_{k}}\left( \mathbf{F} \right) \right),} \nonumber\\ 
		\rm{s.t.}\qquad&{\rm{tr}}\left( {{{\mathbf{\tilde{\Phi }}}}_{k}}\mathbf{F} \right)-{{c}_{k}}\ge \sqrt{2}{{\sigma }_{e,k}}\left\| \mathbf{F} \right\|er{{f}^{-1}}\left( 1-2{{\zeta }_{k}} \right),\forall k, \\ 
		&{\rm{tr}}\left( {{{\mathbf{\overset{\scriptscriptstyle\frown}{\Phi }}}}_{k}}\mathbf{F} \right)-{{\varphi }_{k}}\ge \sqrt{2}{{\beta }_{e,k}}\left\| \mathbf{F} \right\|er{{f}^{-1}}\left( 1-2{{\varepsilon }_{k}} \right),\forall k, \\ 
		&{{\mathbf{F}}_{n,n}}\le 1,\forall n, \\ 
		&\mathbf{F}\succeq 0, \\ 
		&\rm{rank}\left( \mathbf{F} \right)=1. 
	\end{align}
\end{subequations}
Since the constraint (\ref{p2}e) is non-convex, we consider that apply the DC programming to address this non-convex rank-one constraint.

\textbf{Lemma 1:} For any square matrix $\mathbf{B}\in {{\mathbb{C}}^{N\times N}}$, $\mathbf{B}\succeq 0$ and ${\rm{tr}}\left( \mathbf{B} \right)>0$, whose rank is one can be equivalently expressed as
\begin{equation}
	{\rm{rank}}\left( {\bf{{ \mathbf{B}}}} \right) = 1 \Rightarrow {\rm{tr}}\left( {\bf{{ \mathbf{B}}}} \right) - {\left\| {\bf{{ \mathbf{B}}}} \right\|_2}=0,
\end{equation}
where ${\rm{tr}}\left( {\bf{{\mathbf{B}}}} \right) = \sum\limits_{n = 1}^N {{\sigma _n}\left( {\bf{{\mathbf{B}}}} \right)}$, 
${\left\| {\bf{{\mathbf{B}}}} \right\|_2} = {\sigma _1}\left( {\bf{{\mathbf{B}}}} \right)$ represents the spectral norm of matrix $\mathbf{B}$, and represents the $n$-th largest singular value of matrix $\mathbf{B}$.
On the basis of \textbf{Lemma 1}, rank-one constraint (\ref{p2}e) can be transformed in the optimization problem P2 as follows
\begin{equation}\label{rank1}
	{\rm{rank}}\left( {\bf{F}} \right){\rm{ = 1}} \Rightarrow {\rm{tr}}\left( {\bf{F}} \right) - {\left\| {\bf{F}} \right\|_2}=0.
\end{equation}
Then, a penalty factor $\ell $ is introduced  and  the above Eq. (\ref{rank1}) is added to the objective function of the problem P2. Next, it is converted into the problem P3, which can be given by
\begin{subequations}\label{p3}
	\begin{align}
		\text{P3: }\qquad &\underset{\mathbf{F}}{\mathop{\max }}\,\sum\limits_{k=1}^{K}{\left( {{g}_{k}}\left( \mathbf{F} \right)-{{{\bar{g}}}_{k}}{{\left( \mathbf{F} \right)}^{ub}} \right)}-\ell \left( \text{tr}\left( \mathbf{F} \right)-{{\left\| \mathbf{F} \right\|}_{2}} \right),\nonumber\\ 
		\rm{s.t.}\qquad &{\rm{tr}}\left( {{{\mathbf{\tilde{\Phi }}}}_{k}}\mathbf{F} \right)-{{c}_{k}}\ge \sqrt{2}{{\sigma }_{e,k}}\left\| \mathbf{F} \right\|er{{f}^{-1}}\left( 1-2{{\zeta }_{k}} \right),\forall k, \\ 
		& {\rm{tr}}\left( {{{\mathbf{\overset{\scriptscriptstyle\frown}{\Phi }}}}_{k}}\mathbf{F} \right)-{{\varphi }_{k}}\ge \sqrt{2}{{\beta }_{e,k}}\left\| \mathbf{F} \right\|er{{f}^{-1}}\left( 1-2{{\varepsilon }_{k}} \right),\forall k, \\ 
		& {{\mathbf{F}}_{n,n}}\le 1,\forall n, \\ 
		& \mathbf{F}\succeq 0,
	\end{align}
\end{subequations}
where $\ell$ represents the penalty factor associated with the rank-one. Because ${{\left\| \mathbf{F} \right\|}_{2}}$ is a convex function, the problem P3 is still not a convex problem, which can be linearized by using the SCA technique, and its lower bound can be given by
\begin{equation}
	\begin{aligned}
		{{\left\| \mathbf{F} \right\|}_{2}}&\ge {{\left\| {{\mathbf{F}}^{r}} \right\|}_{2}}+{\rm{tr}}\left( {{\mathbf{u}}_{\max }}\left( {{\mathbf{F}}^{r}} \right){{\mathbf{u}}_{\max }}{{\left( {{\mathbf{F}}^{r}} \right)}^{H}}\left( \mathbf{F}-{{\mathbf{F}}^{r}} \right) \right)\\
		&\triangleq {{\left( {{\left\| \mathbf{F} \right\|}_{2}} \right)}^{lb}},
	\end{aligned}
\end{equation}
\newcounter{my13}
\begin{figure*}[!t]
	\normalsize
	\setcounter{my13}{\value{equation}}
	\setcounter{equation}{49}
	\begin{equation}\label{aod}
      \sum\limits_{k = 1}^K {\left( {{{\log }_2}\left( {\sum\limits_{i = 1}^K {{\rho _k}{p_i}{\rm{tr}}\left( {{{\bf{\Phi }}_k}{\bf{F}}} \right){\rm{ + }}{\rho _k}\sigma _k^2 + \delta _k^2} } \right) - {{\log }_2}\left( {{\rho _k}\sum\limits_{i \ne k} {{p_i}{\rm{tr}}\left( {{{\bf{\Phi }}_k}{\bf{F}}} \right) + {\rho _k}\sigma _k^2 + \delta _k^2} } \right)} \right)} = \sum\limits_{k = 1}^K {\left( {{h_k}\left( {{p_i}} \right) - {{\bar h}_k}\left( {{p_i}} \right)} \right).} 
	\end{equation}
	\setcounter{equation}{\value{my13}}
	\hrulefill
	\vspace*{4pt}
\end{figure*}
\newcounter{my13a}
\begin{figure*}[!t]
	\normalsize
	\setcounter{my13a}{\value{equation}}
	\setcounter{equation}{50}
	\begin{equation}\label{aod}
	{\bar h_k}\left( {{p_i}} \right) \le {\bar h_k}\left( {p_i^r} \right) + \frac{{{\rho _k}{\rm{tr}}\left( {{{\bf{\Phi }}_k}{\bf{F}}} \right)}}{{\left( {{\rho _k}\sum\limits_{i \ne k} {p_i^r{\rm{tr}}\left( {{{\bf{\Phi }}_k}{\bf{F}}} \right) + {\rho _k}\sigma _k^2 + \delta _k^2} } \right)\ln 2}}\left( {{p_i} - p_i^r} \right) \buildrel \Delta \over = {\bar h_k}{\left( {{p_i}} \right)^{ub}},\forall k,
	\end{equation}
	\setcounter{equation}{\value{my13a}}
	\hrulefill
	\vspace*{4pt}
\end{figure*}
\newcounter{my14}
\begin{figure*}[!t]
	\normalsize
	\setcounter{my14}{\value{equation}}
	\setcounter{equation}{52}
	\begin{equation}\label{aod}
        \sum\limits_{k = 1}^K {\left( {{{\log }_2}\left( {\sum\limits_{i = 1}^K {{\rho _k}{p_i}{\rm{tr}}\left( {{{\bf{\Phi }}_k}{\bf{F}}} \right){\rm{ + }}{\rho _k}\sigma _k^2 + \delta _k^2} } \right) - {{\log }_2}\left( {{\rho _k}\sum\limits_{i \ne k} {{p_i}{\rm{tr}}\left( {{{\bf{\Phi }}_k}{\bf{F}}} \right) + {\rho _k}\sigma _k^2 + \delta _k^2} } \right)} \right)} =\sum\limits_{k = 1}^K {\left( {f\left( {{\rho _k}} \right) - \bar f\left( {{\rho _k}} \right)} \right),} \forall k. 
	\end{equation}
	\setcounter{equation}{\value{my14}}
	\hrulefill
	\vspace*{4pt}
\end{figure*}where ${{\mathbf{u}}_{\max }}\left( {{\mathbf{F}}^{r}} \right)$ denotes the eigenvector corresponding to the largest eigenvalue of the matrix $\mathbf{F}$ at the $r$-th SCA iteration. Thus, the problem P3 can be further converted
into the problem P4 as follows
\begin{subequations}\label{p4}
	\begin{align}
		\text{P4: }\qquad &\underset{\mathbf{F}}{\mathop{\max }}\,\sum\limits_{k=1}^{K}{\left( {{g}_{k}}\left( \mathbf{F} \right)-{{{\bar{g}}}_{k}}{{\left( \mathbf{F} \right)}^{ub}} \right)}-\ell \left( {\rm{tr}}\left( \mathbf{F} \right)-{{\left( {{\left\| \mathbf{F} \right\|}_{2}} \right)}^{lb}} \right), \nonumber\\ 
		\rm{s.t.}\qquad&{\rm{tr}}\left( {{{\mathbf{\tilde{\Phi }}}}_{k}}\mathbf{F} \right)-{{c}_{k}}\ge \sqrt{2}{{\sigma }_{e,k}}\left\| \mathbf{F} \right\|er{{f}^{-1}}\left( 1-2{{\zeta }_{k}} \right),\forall k, \\ 
		& {\rm{tr}}\left( {{{\mathbf{\overset{\scriptscriptstyle\frown}{\Phi }}}}_{k}}\mathbf{F} \right)-{{\varphi }_{k}}\ge \sqrt{2}{{\beta }_{e,k}}\left\| \mathbf{F} \right\|er{{f}^{-1}}\left( 1-2{{\varepsilon }_{k}} \right),\forall k, \\ 
		& {{\mathbf{F}}_{n,n}}\le 1,\forall n, \\ 
		& \mathbf{F}\succeq 0. 
	\end{align}
\end{subequations}

After the analysis, when the probability of the user's information  and energy outage is less than 0.5, the coefficients on the right side of the inequalities of Eq. (\ref{p4}a) and Eq. (\ref{p4}b) about the matrix $\mathbf{F}$ are positive. In general, the outage probability is not greater than 0.5. The subsequent simulation in this paper is set to 0.1, which can satisfy this condition. If the outage probability is set to be greater than 0.5, SCA can be further used to linearize the right-hand-side (RHS) of the inequalities of Eq. (\ref{p4}a) and Eq. (\ref{p4}b) to solve the problem. Therefore, this problem is a semidefinite programming (SDP) problem, which can be efficiently solved by utilizing the CVX toolbox to obtain the RMS transmissive coefficient.

\subsection{Transmit Power Allocation Optimization}
In this subsection, the RMS transmissive coefficient $\mathbf{F}$ and power splitting ratio $\bm{\rho }$ are given, and we optimize the transmit power allocation $\mathbf{p}$. The objective function can be denoted by Eq. (50). It can be seen that the objective function is the difference of two concave functions w.r.t ${{p}_{i}}$. Thus, it is a non-concave function. It can be linearized by SCA, i.e., we perform a first-order Taylor expansion on the second term and the Eq. (51) can be obtained, where $p_{i}^{r}$ represents the value at the $r$-th SCA iteration. Hence, the problem P1 is transformed as follows
\setcounter{equation}{51}
\begin{subequations}
	\begin{align}
		\text{P5: }\qquad &\underset{\mathbf{p}}{\mathop{\max }}\,\text{ }\sum\limits_{k=1}^{K}{\left( {{h}_{k}}\left( {{p}_{i}} \right)-{{{\bar{h}}}_{k}}{{\left( {{p}_{i}} \right)}^{ub}} \right)}, \nonumber \\ 
		\rm{s.t.}\qquad&{{p}_{k}}\ge 0,\forall k, \\ 
		& \sum\limits_{k=1}^{K}{{{p}_{k}}}\le {{P}_{\max }},\\ 
		& {\rm{tr}}\left( {{{\mathbf{\tilde{\Phi }}}}_{k}}\mathbf{F} \right)-{{c}_{k}}\ge \sqrt{2}{{\sigma }_{e,k}}\left\| \mathbf{F} \right\|er{{f}^{-1}}\left( 1-2{{\zeta }_{k}} \right),\forall k, \\ 
		&{\rm{tr}}\left( {{{\mathbf{\overset{\scriptscriptstyle\frown}{\Phi }}}}_{k}}\mathbf{F} \right)-{{\varphi }_{k}}\ge \sqrt{2}{{\beta }_{e,k}}\left\| \mathbf{F} \right\|er{{f}^{-1}}\left( 1-2{{\varepsilon }_{k}} \right),\forall k. 
	\end{align}
\end{subequations}
Since the problem is a standard convex optimization problem, we can use CVX toolbox to solve it and obtain the transmit power allocation $\mathbf{p}$.
\newcounter{my14a}
\begin{figure*}[!t]
	\normalsize
	\setcounter{my14a}{\value{equation}}
	\setcounter{equation}{53}
	\begin{equation}\label{aod}
	    \bar{f}\left( {{\rho }_{k}} \right) \le \bar{f}\left( \rho _{k}^{r} \right)+\frac{\sum\limits_{i\ne k}{{{p}_{i}}{\rm{tr}}\left( {{\mathbf{\Phi }}_{k}}\mathbf{F} \right)+\sigma _{k}^{2}}}{\left( \rho _{k}^{r}\sum\limits_{i\ne k}{{{p}_{i}}{\rm{tr}}\left( {{\mathbf{\Phi }}_{k}}\mathbf{F} \right)+\rho _{k}^{r}\sigma _{k}^{2}+\delta _{k}^{2}} \right)\ln 2}\left( {{\rho }_{k}}-\rho _{k}^{r} \right)\triangleq \bar{f}{{\left( {{\rho }_{k}} \right)}^{ub}},\forall k.
	\end{equation}
	\setcounter{equation}{\value{my14a}}
	\hrulefill
	\vspace*{4pt}
\end{figure*}

\subsection{Power Splitting Ratio Optimization}
In this subsection, the power splitting ratio $\bm{\rho }$ for each user is optimized when the remaining two sets of variables are fixed. Herein, the objective function can be given by the Eq. (53). Similarly, by using the SCA to linearize the second term in Eq. (53), we can obtain the Eq. (54) on the top of the next page.

Therefore, the problem P1 can be transformed into the problem P6, which can be given by
\setcounter{equation}{54}
\begin{subequations}
	\begin{align}
		\text{P6: }\qquad &\underset{\bm{\rho }}{\mathop{\max }}\,\sum\limits_{k=1}^{K}{\left( f\left( {{\rho }_{k}} \right)-\bar{f}{{\left( {{\rho }_{k}} \right)}^{ub}} \right),} \nonumber\\ 
		\rm{s.t.}\qquad &0\le {{\rho }_{k}}\le 1,\forall k, \\ 
		& {\rm{tr}}\left( {{{\mathbf{\tilde{\Phi }}}}_{k}}\mathbf{F} \right)-{{c}_{k}}\ge \sqrt{2}{{\sigma }_{e,k}}\left\| \mathbf{F} \right\|er{{f}^{-1}}\left( 1-2{{\zeta }_{k}} \right),\forall k, \\ 
		& {\rm{tr}}\left( {{{\mathbf{\overset{\scriptscriptstyle\frown}{\Phi }}}}_{k}}\mathbf{F} \right)-{{\varphi }_{k}}\ge \sqrt{2}{{\beta }_{e,k}}\left\| \mathbf{F} \right\|er{{f}^{-1}}\left( 1-2{{\varepsilon }_{k}} \right),\forall k. 
	\end{align}
\end{subequations}
We can see that this problem is a standard convex optimization problem and can be efficiently solved by utilizing the CVX toolbox.
\subsection{The Overall Robust Joint Optimization Algorithm in Transmissive RMS-enabled SWIPT Networks}
In this subsection, we propose the overall joint RMS transmissive coefficient, transmit power allocation, and power splitting ratio optimization algorithm and summarize it in \textbf{Algorithm 1}. First, when the transmit power allocation and power splitting ratio are given, the RMS transmissive coefficient are determined by solving the problem P4. We can respectively solve the problem P5 and P6 to obtain the transmit power allocation and power splitting ratio. At last, the three sub-problems are optimized alternately until the entire problem converges.
\begin{algorithm}[H]
	\caption{Robust Joint Optimization Algorithm in Transmissive RMS-enabled SWIPT Networks } 
	\begin{algorithmic}[1]
		\State$\textbf{Input:}$ ${{\bf{F}}^0}$, ${{\bf{p}}^0}$, ${{\bm{\rho}}^0}$, threshold $\epsilon$ and iteration index $r = 0$.
		\Repeat
		\State Solve the problem P4 to obtain RMS transmissive coefficient ${\bf{F}}^ * $.
		\State Solve the problem P5 to obtain transmit power allocation ${\bf{p}}^ * $.
		\State Solve the problem P6 to obtain power splitting ratio ${\bm{\rho}}^ * $.
		\State Update iteration index  $r=r+1$.
		\Until The whole problem satisfies the convergence threshold requirement.
		\State \Return RMS transmissive coefficient, transmit power allocation, power splitting ratio.
	\end{algorithmic}
\end{algorithm}
\subsection{Computational Complexity and Convergence Analysis}
\subsubsection{Computational complexity analysis}
In each iteration, the computational complexity of the proposed robust joint optimization algorithm is divided into three parts. The first is to solve the SDP problem P4 with complexity ${\cal O}\left( {{M^{3.5}}} \right)$ through the interior point method \cite{boyd2004convex}. In addition, the complexity of calculating the subgradient through singular value decomposition is ${\cal O}\left( {{M^3}} \right)$. Accordingly, the computational complexity of the first part is at most ${\cal O}\left( {{M^{3.5}}} \right)$. Then, both the second part and the third part solve problems P5 and P6 with computational complexity ${\cal O}\left( {{K^{3.5}}} \right)$, respectively. Herein, let $r$ be the number of iterations required for the proposed robust joint optimization algorithm to reach convergence, the computational complexity of \textbf{Algorithm 1} can be expressed as ${\cal O}\left( {r\left( {{K^{3.5}} + {M^{3.5}}} \right)} \right)$.

\subsubsection{Convergence analysis}
The convergence of the proposed robust joint optimization \textbf{Algorithm 1} in transmissive RMS-enabled SWIPT networks can be proved as as shown later.

Let ${{\bf{F}}^r}$, ${{\bf{p}}^r}$ and ${{\bm{\rho}}^r}$ denote the $r$-th iteration solution to the problem P4, P5 and P6. The objective function can be expressed as ${\cal R}\left( {{{\bf{F}}^r},{{\bf{p}}^r},{{\bm{\rho }}^r}} \right)$. In the step 3 of \textbf{Algorithm 1}, the RMS transmissive coefficient ${\bf{F}}^ * $ can be obtained for given ${{\bf{p}}^r}$ and ${{\bm{\rho}}^r}$. Hence, we have 
\begin{equation}
	{\cal R}\left( {{{\bf{F}}^r},{{\bf{p}}^r},{{\bm{\rho }}^r}} \right) \le {\cal R}\left( {{{\bf{F}}^{r + 1}},{{\bf{p}}^r},{{\bm{\rho }}^r}} \right).
\end{equation}
In the step 4 of \textbf{Algorithm 1}, the transmit power allocation ${\bf{p}}^ * $ can be obtained when ${{\bf{F}}^r}$ and ${{\bm{\rho}}^r}$ are given. Herein, we also have 
\begin{equation}
	{\cal R}\left( {{{\bf{F}}^{r + 1}},{{\bf{p}}^r},{{\bm{\rho }}^r}} \right) \le {\cal R}\left( {{{\bf{F}}^{r + 1}},{{\bf{p}}^{r + 1}},{{\bm{\rho }}^r}} \right).
\end{equation}
Similarly, in the step 5 of \textbf{Algorithm 1}, the power splitting ratio ${\bm{\rho}}^ * $ can also be obtained when ${{\bf{F}}^r}$ and ${{\bf{p}}^r}$ are given. Thus, we have
\begin{equation}
	{\cal R}\left( {{{\bf{F}}^{r + 1}},{{\bf{p}}^{r + 1}},{{\bm{\rho }}^r}} \right) \le {\cal R}\left( {{{\bf{F}}^{r + 1}},{{\bf{p}}^{r + 1}},{{\bm{\rho }}^{r + 1}}} \right).
\end{equation}
Based on the above, we can obtain
\begin{equation}
	{\cal R}\left( {{{\bf{F}}^r},{{\bf{p}}^r},{{\bm{\rho }}^r}} \right) \le {\cal R}\left( {{{\bf{F}}^{r + 1}},{{\bf{p}}^{r + 1}},{{\bm{\rho }}^{r + 1}}} \right).
\end{equation}
The above inequality proves that the value of the objective function  is monotonic non-decreasing after each iteration of \textbf{Algorithm 1}. In addition, there is an upper bound on the objective function value for the problem P1. The above two aspects ensure the convergence performance of \textbf{Algorithm 1}.

\section{Numerical Results}
In this section, we demonstrate the effectiveness of the proposed robust joint optimization algorithm in transmissive RMS-enabled SWIPT networks through numerical simulations. In the simulation setting, we consider a three-dimensional communication network scenario, where the position of RMS transmitter is (0m, 0m, 15m), and $K=4$ users are randomly distributed in a circle whose center coordinates is (0m, 0m, 0m) with a radius of 50m. RMS is equipped with $N=16$ elements. The antenna spacing is set to half the wavelength of the carrier. Meanwhile, we assume that the parameters of all users are the same, i.e.,  ${a_k} = 150$, ${b_k} = 0.024$ and ${\xi _k} = 24{\rm{mW}}$\cite{8478252}. Herein, we set $\sigma ^2=-50$dBm, ${\gamma}_{th} =-30$dB and ${E}_{th} =-40$dB in the simulations. The path loss exponent is set as $\alpha=3$. We set the path loss $\beta$ to -20dB when the reference distance is 1m and set the Rician factor $\kappa$ to 3dB. The threshold for algorithm convergence is set as $10^{-3}$.

First, the convergence of the proposed algorithm is verified in transmissive RMS-enabled SWIPT networks. Fig. 2 shows the change of system sum-rate with algorithm iterations. It obvious that the sum-rate increases as the number of iterations increases, which verifies our proposed algorithm has good convergence. In addition, we compare the effect of different RMS transmissive element counts on system performance. Considering that the array of RMS is distributed in a UPA with the same number of elements in the horizontal and vertical directions, the number of RMS element is a perfect square. Specifically, we compare the system sum-rate of the proposed algorithm when the number of RMS transmissive elements are 9, 16, and 36. It can be concluded that the larger the number of RMS elements, the upper the system sum-rate.
\begin{figure}
	\centerline{\includegraphics[width=9.5cm]{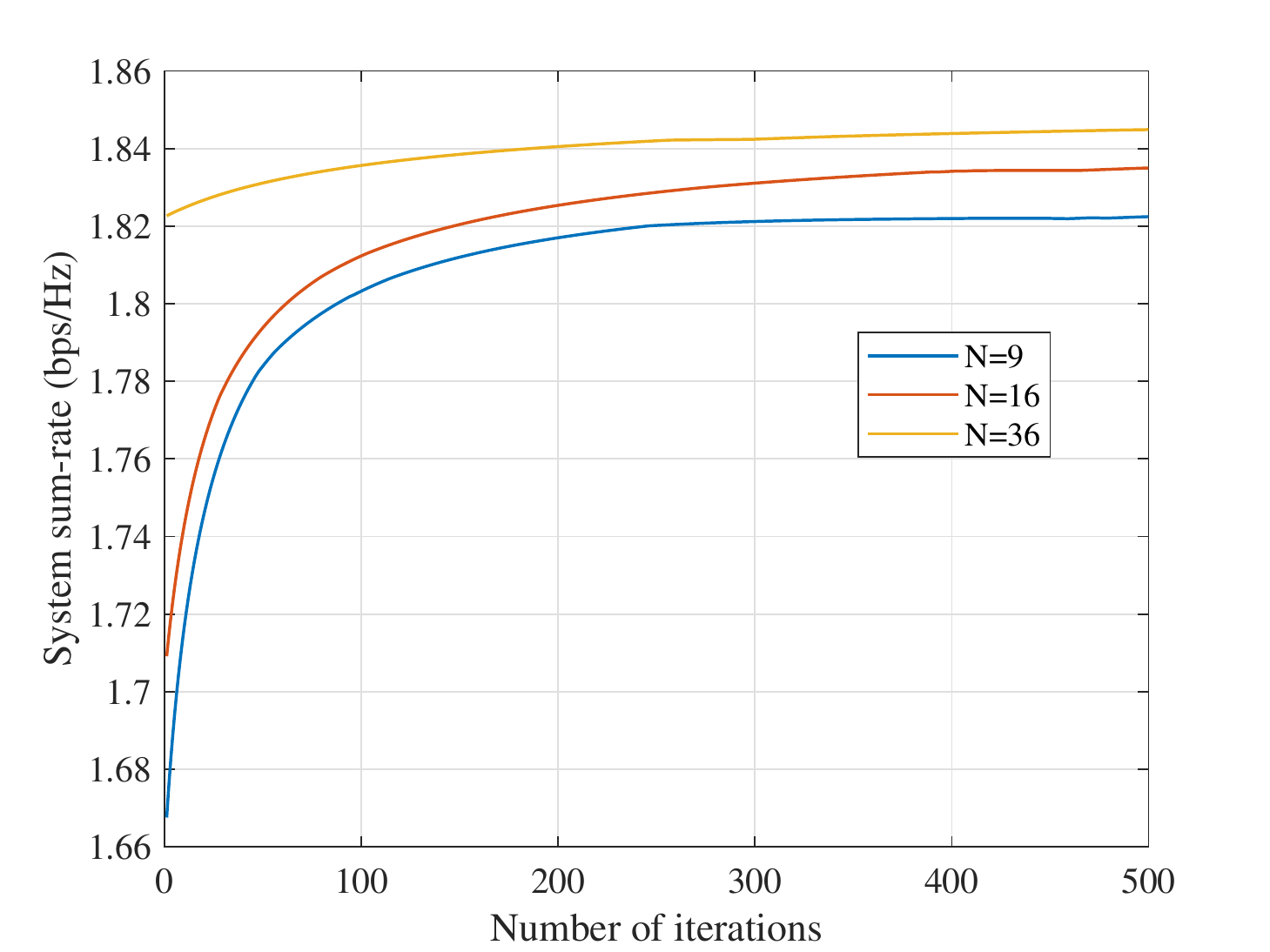}}
	\caption{Convergence behavior of the proposed robust joint optimization algorithm.}
\end{figure}

In this section, We verify the good performance of the proposed robust joint optimization algorithm 
in transmissive RMS-enabled SWIPT networks
compared with other benchmark algorithms. (1) benchmark 1 (RMS-random-phase): In this case, we adopt a random RMS coefficient to deploy RMS and don't optimize its coefficient, i.e., the problems P4 and P6 are optimized alternately. (2) benchmark 2 (fixed-transmit-power): In this case, the transmit power is allocated equally to each user, while the RMS transmissive coefficient and power splitting ratio optimization still use the solution of problems P4 and P6. (3) benchmark 3 (fixed-power-splitting-ratio): In this case, the power splitting ratio is regarded as a constant, i.e., ${\rho _k} = 0.5,\forall k$, and we optimize problems P4 and P5 jointly.

Next, we investigate the relationship between the system sum-rate and the maximum transmit power of RMS transceiver. As shown in Fig. 3, the system sum-rate increases as the increase of the maximum transmit power of RMS transceiver, and the performance of our proposed algorithm outperforms all benchmarks, which reflects the advantage of jointly optimizing the RMS transmissive coefficient, transmit power allocation and power splitting ratio. The performance of benchmark 2 is the worst because it equally allocates the transmit power to each user and does not take advantage of the channel differences of different users. The performance of the system can be improved by allocating more resources to users whose channel quality is better. Compared with benchmark 3, the proposed scheme has similar performance when the transmit power is high, because when the power is high, the constraints of the user's SINR and energy harvested are easier to meet, and the system performance mainly depends on the transmissive RMS coefficient design and transmit power allocation.
\begin{figure}
	\centerline{\includegraphics[width=9.5cm]{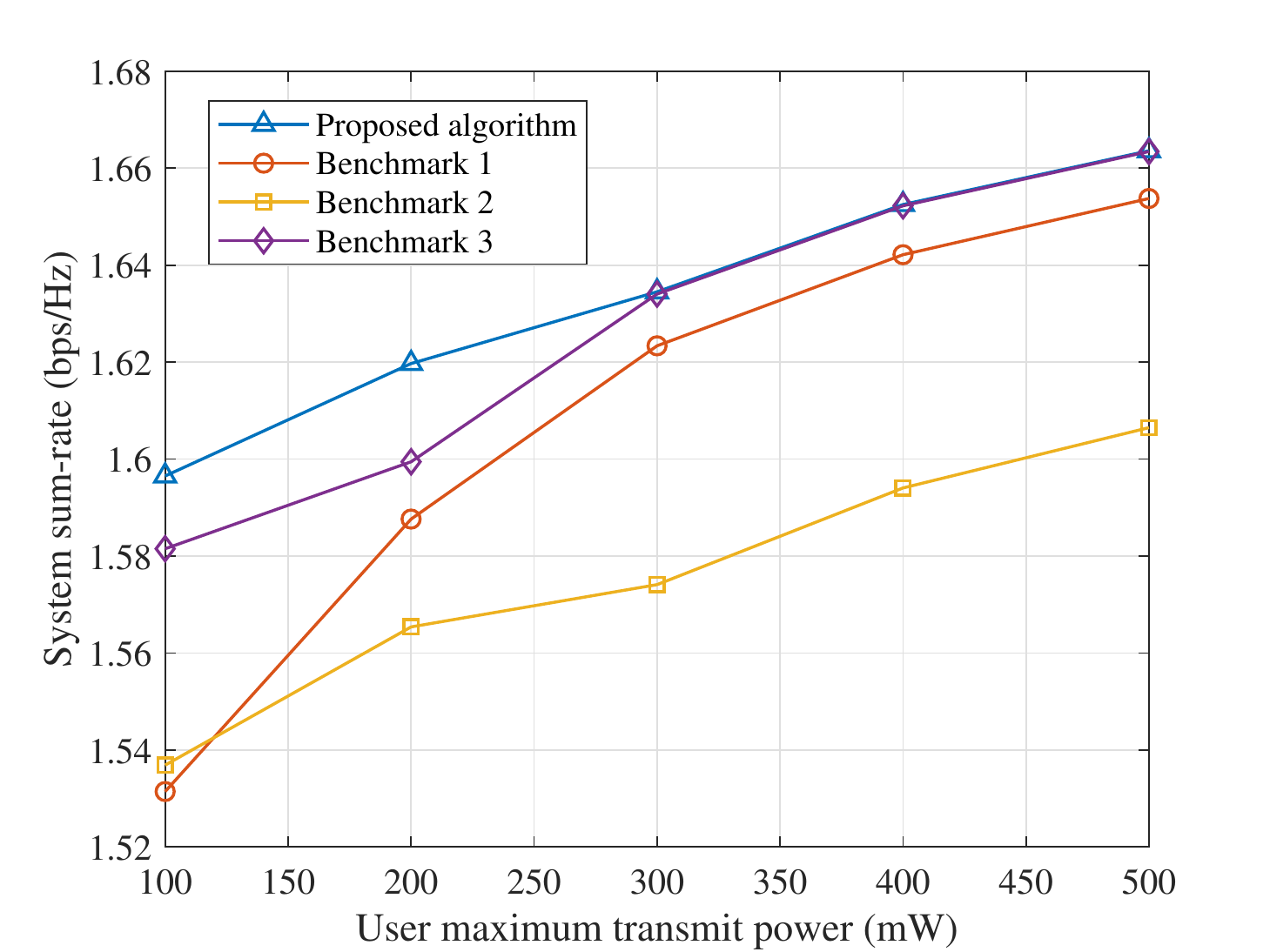}}
	\caption{System sum-rate versus maximum transmit power.}
\end{figure}

Fig. 4 shows the system sum-rate verus the number of RMS transmissive elements. It can be seen that the system sum-rate increases as the number of transmissive RMS element increases for all benchmark algorithms, which
is mainly because when the number of transmissive elements increases, the number of reconstructed channels increases, and the channel gain of the receiver increases. This also reflects the performance advantage of the RMS as a low-cost passive component, which improves spatial diversity by increasing the number of RMS elements without requiring additional signal processing. It has a wide range of application in IoT networks. Moreover, the proposed algorithm has obvious performance advantages in different numbers of RMS elements, which reflecting the advantage of the robust joint optimization algorithm.
\begin{figure}
	\centerline{\includegraphics[width=9.5cm]{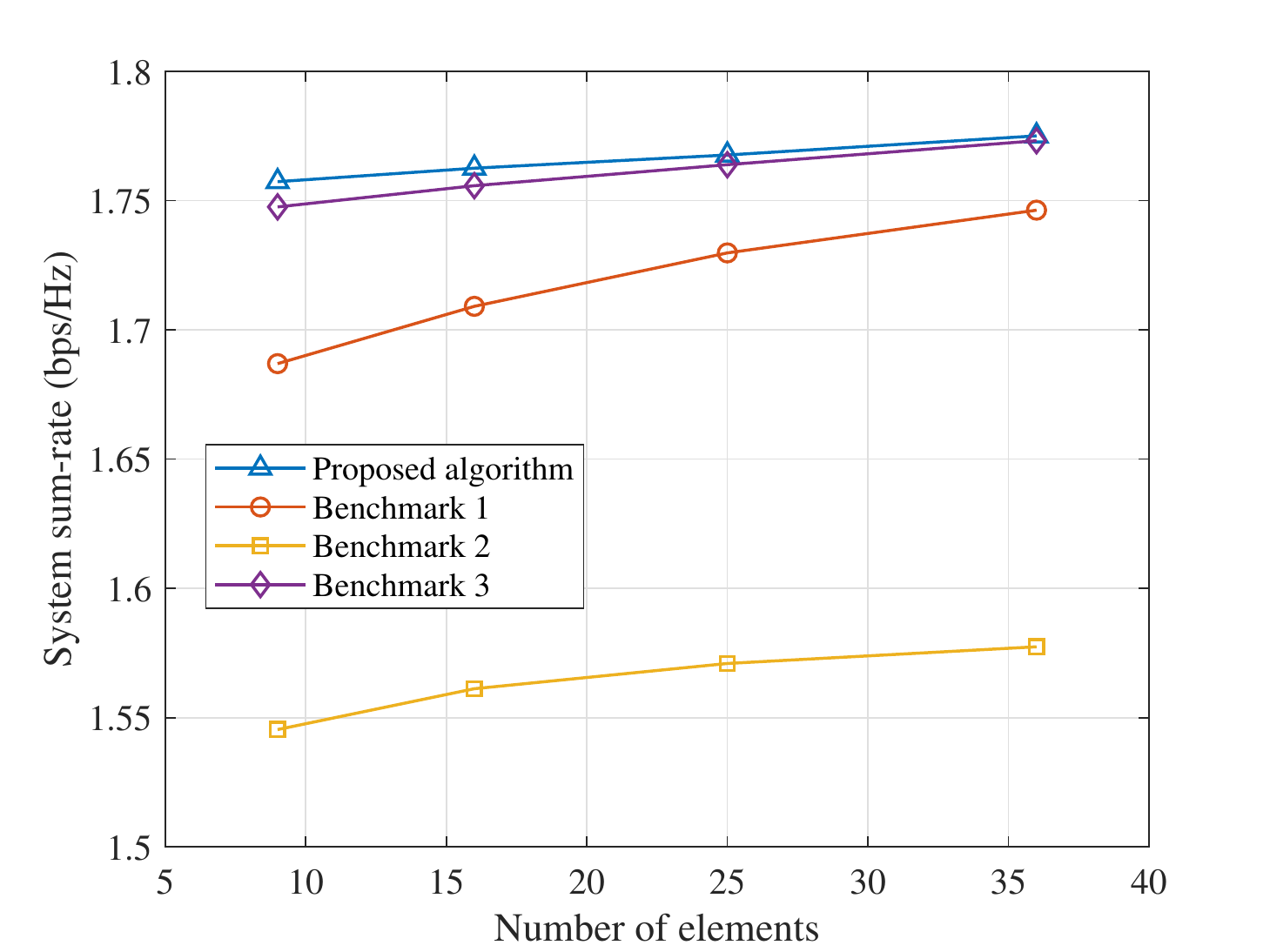}}
	\caption{System sum-rate versus the number of RMS transmissive elements.}
\end{figure}

Then, the system sum-rate versus the number of users is dipicted in Fig. 5. It is obvious that the system sum-rate decreases as the increase of the number of users. This is mainly because we keep the maximum transmit power unchanged. When the number of users increases and the SINR constraints of each user must be satisfied, each user needs to obtain a certain amount of energy, which leads to mutual interference increases and users with better channels have difficulty obtaining more power. Furthermore, our proposed algorithm still outperforms other benchmarks with the same number of users, which indicates our proposed algorithm can better deal with mutual interference.
\begin{figure}
	\centerline{\includegraphics[width=9.5cm]{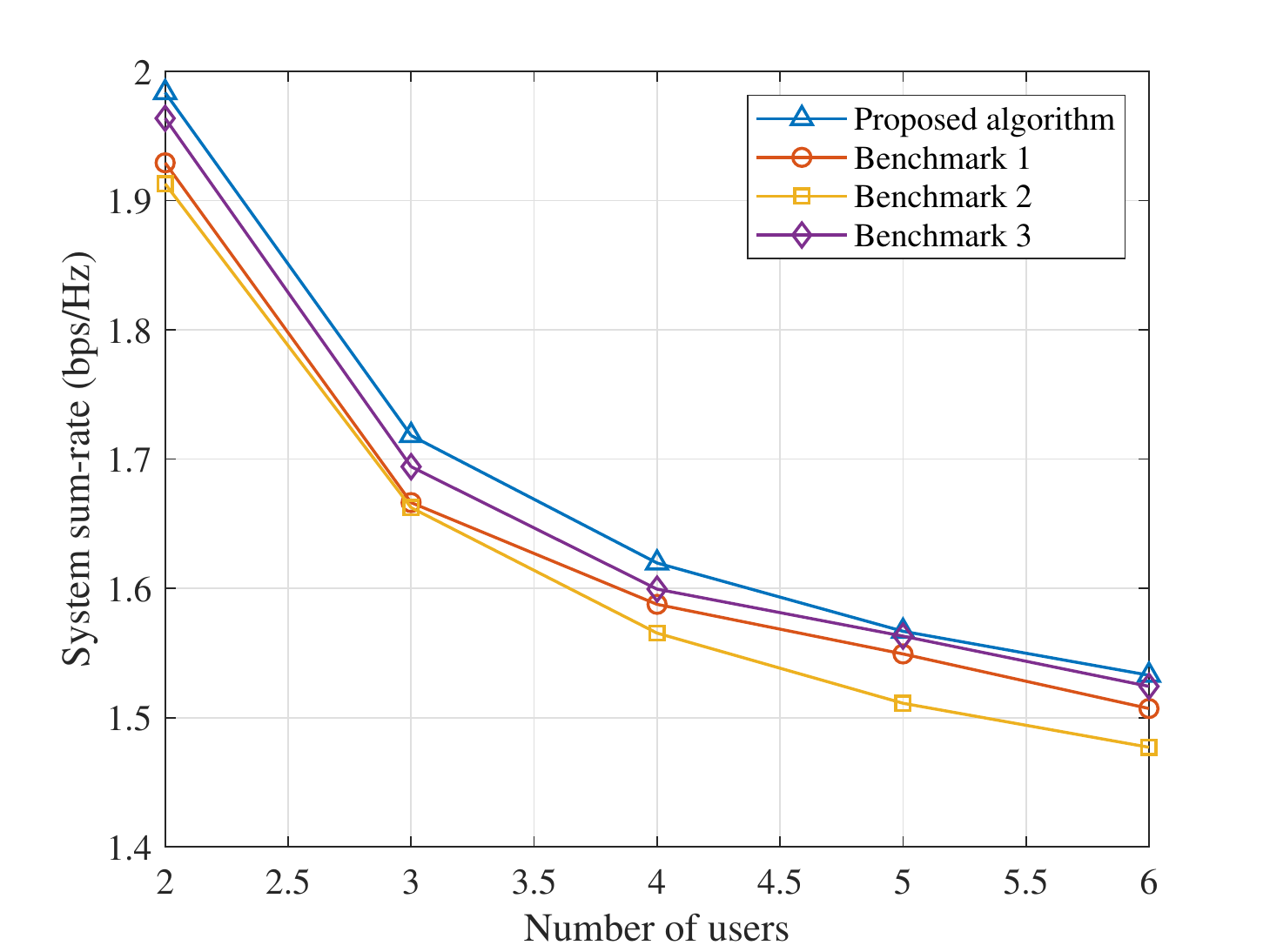}}
	\caption{System sum-rate versus the number of users.}
\end{figure}

Fig. 6 shows the system sum-rate versus energy harvested threshold. It is obvious that when the user's energy harvested threshold increases, the system sum-rate decreases. Owing to when the threshold increases, the user needs to obtain a larger power allocation or decrease the power splitting ratio to meet the constraints of energy harvested, and the achievable rate of each user decreases with the decrease of the power splitting ratio. Therefore, system sum-rate also decreases. While the performance of the benchmark 3 remains almost unchanged, this is because we satisfy the constraints by initially setting a reasonable splitting ratio, and the power splitting ratio will not change in the subsequent alternate optimization process. 
\begin{figure}
	\centerline{\includegraphics[width=9.5cm]{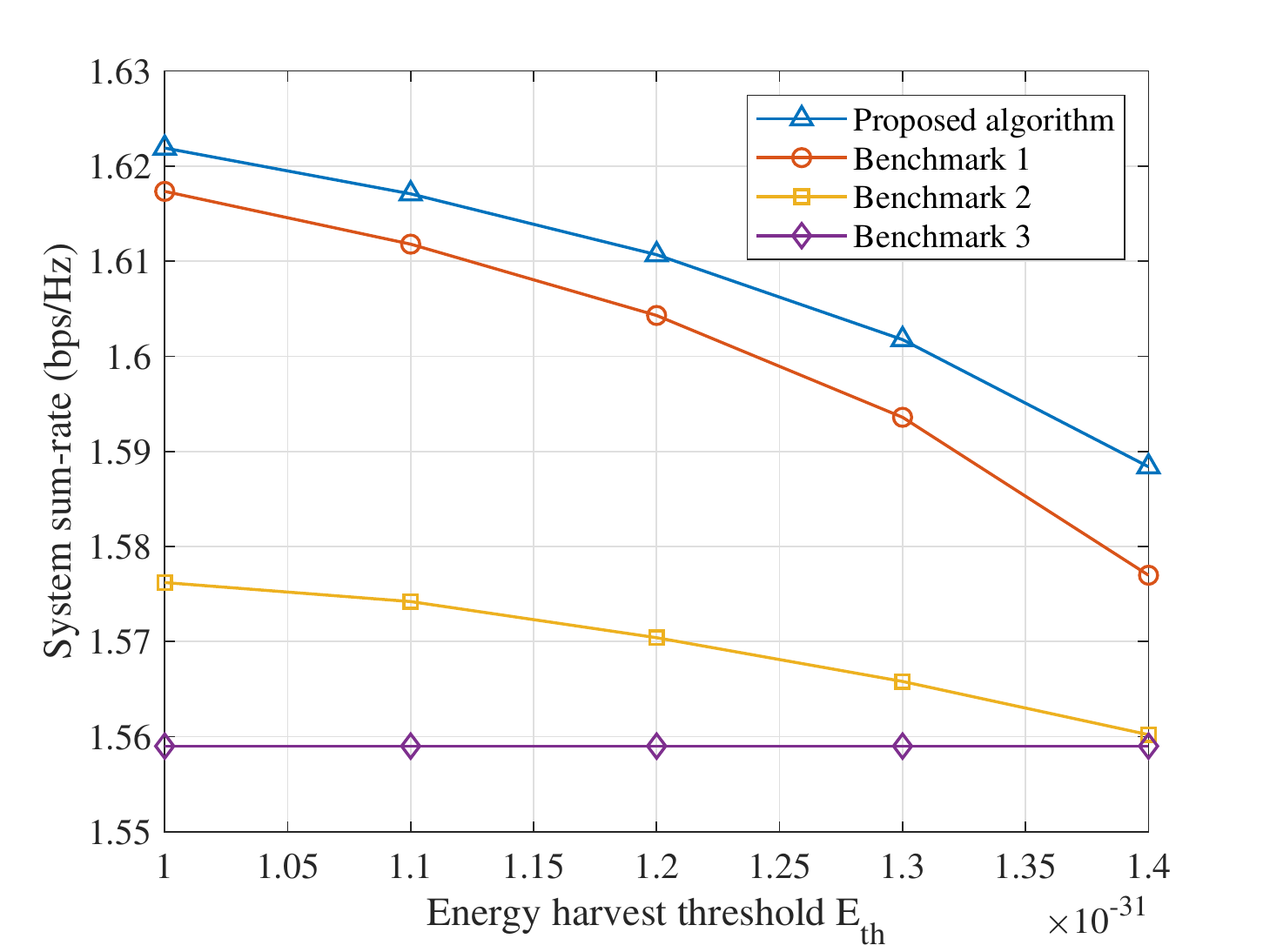}}
	\caption{System sum-rate versus the energy harvested threshold.}
\end{figure}

Fig. 7 depicts the system sum-rate versus the noise power spectral density. It can be seen that the performance is greatly affected by noise due to the large interference between users in the model considered in this paper. As the noise power spectral density increases, the system sum-rate decreases. Compared with other benchmarks, our proposed algorithm has the best performance, and the advantage is more obvious in the environment with larger noise, which shows that our proposed optimization of joint RMS transmissive coefficient, transmit power allocation and power splitting ratio can be used well in all environments.
\begin{figure}
	\centerline{\includegraphics[width=9.5cm]{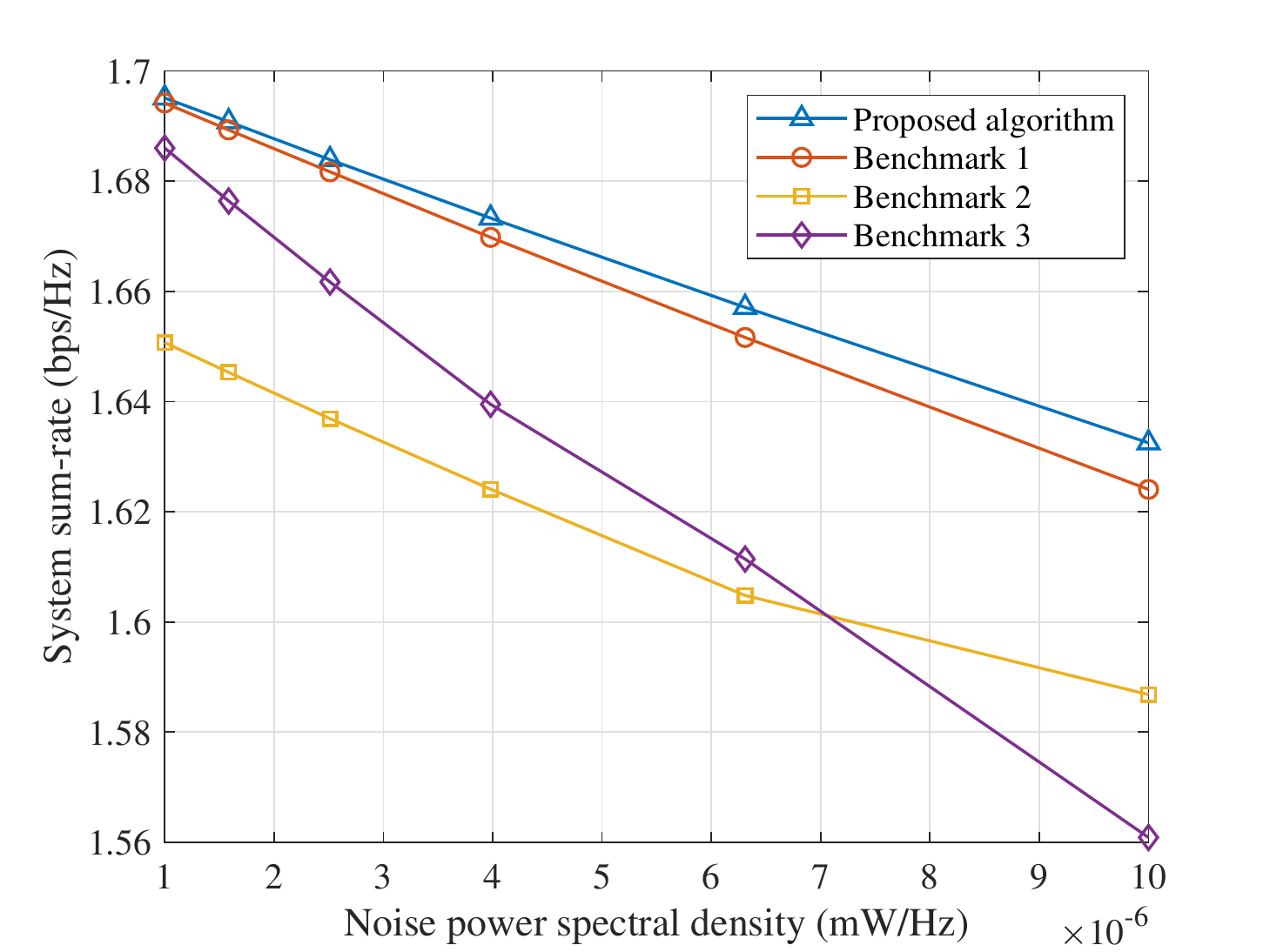}}
	\caption{System sum-rate versus the noise power spectral density.}
\end{figure}

Fig. 8 illustrates the variation of our proposed robust joint optimization algorithm and other benchmark algorithms versus different spectral norm of channel error matrix. The abscissa of Fig. 8 is logarithmic. The increase of channel estimation error will lead to the degradation of system performance. This is mainly because a larger channel estimation error matrix will make the constraints (22d) and (22e) tighter, which will degrade performance of the system in terms of sum-rate. It's obvious that when the spectral norm of channel error matrix is large, the performance of benchmark 3 decreases sharply. This is because a small power splitting ratio is set to meet the requirements of constraint (22e) initially, and power splitting ratio cannot be updated during the alternate optimization process, and the objective function is significantly affected by power splitting ratio at this time. In fact, perfect CSI cannot be obtained at the transmitter in the practical system, a certain channel estimation error is considered in our model, which is more robust and more conducive for deployment in actual communication networks.
\begin{figure}
	\centerline{\includegraphics[width=9.5cm]{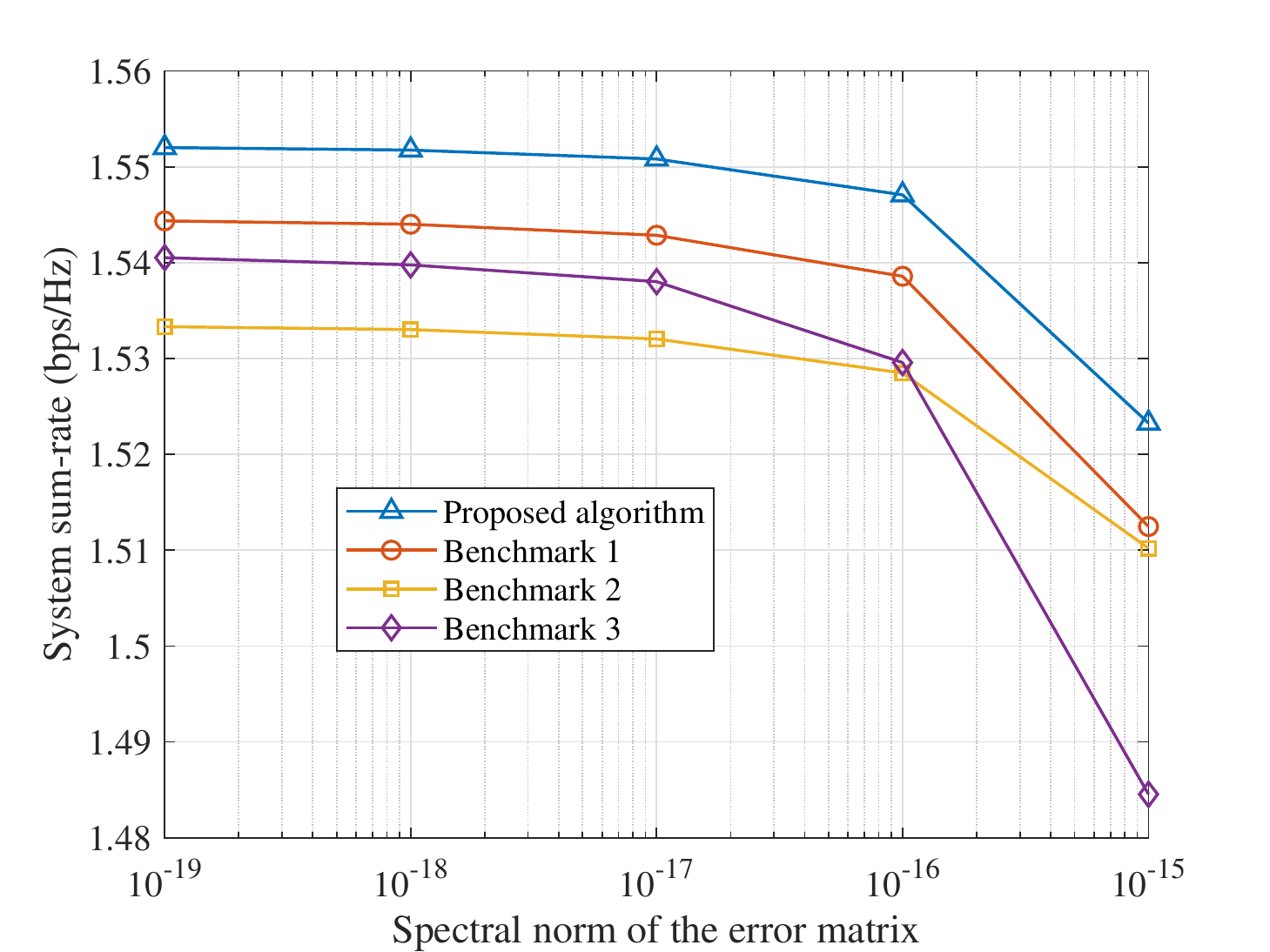}}
	\caption{System sum-rate versus spectral norm of channel error matrix.}
\end{figure}

\section{Conclusions}
In this paper, we investigate the system sum-rate maximization problem for transmissive RMS-enabled SWIPT networks. Specifically, RMS transmissive coefficient, transmit power allocation and power splitting ratio are jointly designed under the requirements of SINR and energy harvested based on outage probability criterion. First, the problem containing outage probability constraints is transformed into a tractable optimization problem. Owing to non-convexity of the transformed problem, AO algorithm based on SCA, DC and penalty function method is implemented to to handle non-convexity and solve the problem. Besides, we analyze the complexity of the proposed algorithm and prove its convergence performance. From the numerical results, it can be concluded that our proposed algorithm outperforms other algorithms in terms of system sum-rate, which demonstrate transmissive RMS transceiver is a potential multi-antenna technology in the design of future wireless communication networks.

\bibliographystyle{IEEEtran}
\bibliography{reference}

\end{document}